# Universal Minimax Discrete Denoising under Channel Uncertainty[*]

George Gemelos   Styrmir Sigurjónsson   Tsachy Weissman

October 15, 2018

**Abstract**

The goal of a denoising algorithm is to recover a signal from its noise-corrupted observations. Perfect recovery is seldom possible and performance is measured under a given single-letter fidelity criterion. For discrete signals corrupted by a *known* discrete memoryless channel, the DUDE was recently shown to perform this task asymptotically optimally, without knowledge of the statistical properties of the source. In the present work we address the scenario where, in addition to the lack of knowledge of the source statistics, there is also uncertainty in the channel characteristics. We propose a family of discrete denoisers and establish their asymptotic optimality under a minimax performance criterion which we argue is appropriate for this setting. As we show elsewhere, the proposed schemes can also be implemented computationally efficiently.

## 1 Introduction

Discrete sources corrupted by Discrete Memoryless Channels (DMCs) are encountered naturally in many fields, including information theory, computer science, and biology. The reader is referred to [15] for examples, as well as references to some of the related literature. It was shown in [15] that optimum denoising of a finite-alphabet source corrupted by a *known* invertible[1] DMC can be achieved asymptotically, in the size of the data, without knowledge of the source statistics. It was further shown that the scheme achieving this performance, the Discrete Universal DEnoiser (DUDE), enjoys properties that are desirable from a computational view point.

The assumption of a known channel in the setting of [15] is integral to the construction of the DUDE algorithm. This assumption is indeed a realistic one in many practical scenarios where the noisy medium through which the data is acquired is well characterized statistically. Furthermore, the computational simplicity of the DUDE allows it to be used in certain cases when the statistical properties of the DMC may not be fully known. For example, when there is a human observer to give feedback on the quality of the reconstruction. In such a case, the human observer can scan through the various possible DMCs, implementing the DUDE for each DMC, and select the one which gives the best reconstruction. Such a method can be used to extend the scheme of [15] to the case of channel uncertainty when it is reasonable to expect the availability of feedback on the quality of the reconstruction.

---

[*]Authors are with the department of electrical engineering, Stanford University, Stanford, CA 94305. Email: ggemelos@stanford.edu, styrmir@stanford.edu, and tsachy@stanford.edu.

The work of the first two authors was supported by MURI Grant DAAD-19-99-1-0215 and NSF Grants CCR-0311633 and CCF-0512140. The work of the third author was supported in part by NSF Grant CCR-0312839.

[1]Throughout this paper, "invertible DMC" is one whose associated channel matrix is of full row rank.

Unfortunately, such feedback is not realistic in many scenarios. For example, in applications involving DNA data [16], a human observer would probably find the task of determining which two reconstructions of a corrupted nucleotide sequence is closer to the original quite difficult. Other examples include applications involving the processing of large databases of noisy images [9] and those involving medical images [17]. In the latter, human feedback is often too subjective. In such cases, an automated algorithm for discrete image denoising which can accommodate uncertainty in the statistical characteristics of the noisy medium is desired. With this motivation in mind, in this paper we address the problem of denoising when, in addition to the lack of knowledge of the source statistics, there is also uncertainty in the channel characteristics.

It turns out that the introduction of uncertainty in the channel characteristics into the setting of [15] results in a fundamentally different problem, calling for new performance criteria and denoising schemes which are principally different than those of [15]. The main reason for this divergence is that in the presence of channel uncertainty, the distribution of the noise-corrupted signal does *not* uniquely determine the distribution of the underlying clean signal, a property which is key to the DUDE of [15] and its accompanying performance guarantees. To illustrate this difference, consider the simple example of the Bernoulli source corrupted by a Binary Symmetric Channel (BSC). In this example, the noise-corrupted signal is also Bernoulli with some parameter $\delta < 1/2$. For simplicity, we will only consider two possibilities: either the clean signal is the "all zero" signal corrupted by a BSC with crossover probability $\delta$ or the clean signal is Bernoulli($\delta$) passed through a noise-free channel.[2] It is easy to see that solely knowing that the noise-corrupted signal is Bernoulli($\delta$), there is no way to distinguish between the two possibilities above. It is therefore impossible to uniquely identify the distribution of the underlying source. Degenerate as this example may be, it highlights the following points, which are key to our present setting and its basic difference from that of [15]:

1. Even with complete knowledge of the noise-corrupted signal statistics, Bernoulli($\delta$) in our example, there is no way of inferring the distribution of the underlying source.

2. There exists no denoising scheme that is simultaneously optimal for all, two in our example, sources which can give rise to the noise-corrupted signal statistics.

3. A scheme that minimizes the worst case loss has to be randomized.[3] In the example above, the scheme that minimizes the worst case bit error rate is readily seen to be the one which randomizes, equiprobably, between using the observed noisy symbol as the estimate of the clean symbol and estimating with the 0 symbol regardless of the observation. Such a scheme would achieve a bit error rate of $\delta/2$ under both possible sources discussed above.

As is evident through this example, the key issue is that while in the setting of [15] there is a one-to-one correspondence between the channel output distribution and its input distribution, a channel output distribution can correspond to many input distributions in the presence of channel uncertainty. This point has also been a central theme in [4, 12], where fundamental performance limits are characterized for rate constrained denoising under uncertainty in both the source and channel characteristics.[4]

---
[2]Throughout this paper, Bernoulli($\delta$) refers to a Bernoulli process with parameter $\delta$.
[3]Either in "space" (i.e., true randomization) or in time (i.e., time sharing for deterministic estimates.)

Under these circumstances, given any noise-corrupted signal, a seemingly natural criterion under which the performance of a denoising scheme should be judged is its worst case performance under all source-channel pairs that can give rise to the observed noise-corrupted signal statistics. In line with this conclusion, as a way to evaluate the merits of a denoising scheme, we look at a scheme's worst case performance assessed by a third party that has complete knowledge of both the noise-corrupted signal distribution *and* the whole noise-corrupted signal realization. Under this criterion, we define the notion of "sliding window denoisability" to be the best performance attainable by a sliding window scheme of any order. This can be considered our setting's analogue to the "sliding window denoisability" of [15] (which in turn was inspired by the finite-state compressibility of [18], the finite-state predictability of [7], and the finite-state noisy predictability of [14]). By definition, this is a fundamental lower bound on the performance of any sliding window scheme. Our main contribution is the presentation of a family of sliding window denoisers that asymptotically attains this lower bound.

The problem of denoising discrete sources corrupted by an unknown DMC has been previously considered in the context of state estimation in the literature on hidden Markov models (cf. [6] and the many references therein). In that setting, one assumes the source to be a Markov process. The EM algorithm of [2] is then used to obtain the maximum likelihood estimates of the process and channel parameters. One then denoises optimally assuming the estimated values of the source and channel parameters. This approach is widely employed in practice and has been quite successful in a variety of applications. Other than the hidden Markov model method, the only other general approach we are aware of for discrete denoising under channel uncertainty is the DUDE with "feedback" discussed above. For the special case of binary signals corrupted by a BSC, an additional scheme was suggested in [15, Subsection 8-C] which makes use of a particular estimate of the channel crossover probability.

These existing schemes lack solid theoretical performance guarantees. Insofar as the hidden Markov model based schemes go, performance guarantees are available only for the case where the underlying source is a Markov process. Furthermore, these performance guarantees stipulate "identifiability" conditions (cf. [1, 11] and references therein), which do not hold in our setting of channel uncertainty. The more recent approach of employing the DUDE tailored to an estimate of the channel characteristics is shown in [8] to be suboptimal with respect to the worst case performance criterion we propose. This suggests that the schemes we introduce in this work are of an essentially different nature than the DUDE [15].

After we state the problem in Section 2, we turn to describe our denoiser in Section 3. In Section 4 we concretely introduce the performance measure and performance benchmarks that were qualitatively described above for the case where there are a finite number of possible channels. In Section 5 we state our main results, which assess the performance of the denoisers of Section 3 and guarantee their universal asymptotic optimality under the performance criteria of Section 4. To focus on the essentials of the problem, we assume in Section 5 that the channel uncertainty set is finite. In Section 6, we extend the performance measure of Section 4 and the guarantees of Section 5 to the case of an infinite number of possible channels. The proof of the results are left to the appendix.

---

[4]In that line of work, Shannon theoretic aspects of the problem are considered and attention is restricted to memoryless sources. Our current framework considers noise-free sources that are arbitrarily distributed.

## 2 Problem Statement

Before formally stating the problem, we introduce some notation: An upper case letter will denote random quantities while the corresponding lower case letter will denote individual realizations. Bold notation will be used to represent doubly infinite sequences. For example, $\mathbf{X}$ will denote the stochastic process $\{\ldots, X_{-1}, X_0, X_1, \ldots\}$ and $\mathbf{x} = \{\ldots, x_{-1}, x_0, x_1, \ldots\}$ a particular realization. Furthermore, for indices $i \leq j$, the vector $(X_i, \ldots, X_j)$ will be denoted by $X_i^j$. We will omit the subscript when $i = 1$.

Using the above notation, the problem statement is as follows: Let $\Delta$ be a collection of invertible DMCs. A source $\mathbf{X}$ is passed through an unknown DMC in $\Delta$ and we denote the output process as $\mathbf{Z}$. The process $\mathbf{Z}$ is thus a noise-corrupted version of the $\mathbf{X}$ process. We assume that the components of both $\mathbf{X}$ and $\mathbf{Z}$ take on values in a finite alphabet denoted by $\mathcal{A}$. Given $Z^n$ and $\Delta$, we wish to reconstruct $X^n$ under a given single letter loss function, $\Lambda : \mathcal{A} \times \mathcal{A} \mapsto \mathbb{R}^+$. For $a, b \in \mathcal{A}$, $\Lambda(a, b)$ can be interpreted as the loss incurred when reconstructing the symbol $a$ with the symbol $b$. Here we make the assumption that the components of the reconstruction also lie in the finite alphabet $\mathcal{A}$. Given $x^n, \hat{x}^n \in \mathcal{A}^n$, we denote

$$\Lambda(x^n, \hat{x}^n) = \frac{1}{n} \sum_{i=1}^{n} \Lambda(x_i, \hat{x}_i).$$

## 3 Description of the Algorithm

Inherent in the setup of our problem is the uncertainty regarding which channel corrupted the clean source, as depicted in Figure 1. We are given that the channel lies in an uncertainty set $\Delta$, and the uncertainty set is assumed to be fixed and known to the denoiser. The description of the denoiser is broken into two parts. In Section 3A we present an overview of the development of the denoiser, while a detailed construction of the denoiser is presented in Section 3B.

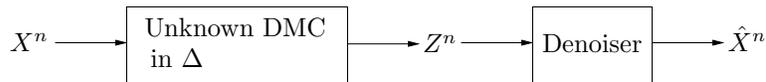

Figure 1: A noiseless source $X^n$, corrupted by a channel known to lie in an uncertainty set $\Delta$, and we observe the output $Z^n$.

### A Outline of Algorithm

For simplicity, we start by limiting $\Delta$ to be a finite collection of invertible DMCs. The case of $|\Delta|$ being infinite requires a more technical analysis which will be discussed in Section 6. Throughout the paper, we confine our discussion to *sliding window* denoisers. A sliding window denoiser of order $k$ works as follows: When denoising a particular symbol, it considers the $k$ symbols preceding it and the $k$ symbols succeeding it. These $k$ symbols before and after the current symbol form a two-sided context of the current symbol. In particular, if we denote the current symbol by $z_0$, the two-sided context is $z_{-k}^{-1}$ and $z_1^k$. In addition to the usual deterministic denoisers, we allow randomized denoisers. A randomized denoiser is a denoiser whose output is a distribution from which a reconstruction must be drawn as a final step. Therefore, we can think of a sliding window denoiser, both deterministic and random, as a mapping from $\mathcal{A}^{2k+1} \mapsto \mathcal{S}(\mathcal{A})$. Here, for a given alphabet $\mathcal{A}$, $\mathcal{S}(\mathcal{A})$ is used to denote the $|\mathcal{A}|$-dimensional probability simplex[5]. If $f$ is a sliding window denoiser, we denote its simplex-valued output by $f([z_{-k}^{-1}, z_0, z_1^k])$ or $f(z_{-k}^k)$. We can use a $k$-order sliding window denoiser $f$ to denoise $Z^n$, by drawing the $i$-th reconstruction according to the distribution $f(Z_{i-k}^{i+k})$.

Let $\Pi$ be some channel in $\Delta$, $P_{Z_{-k}^k}$ the probability distribution on $Z_{-k}^k$, and $f$ a sliding window denoiser.[6] We now assume there exists a function $G_k$ that, when given $\Pi$, $P_{Z_{-k}^k}$, and $f$, evaluates the performance of the denoiser $f$ on that particular $\Pi$ and $P_{Z_{-k}^k}$. Here performance is measured by the expected loss, under $\Lambda$, incurred when estimating $X_0$ based on $f(Z_{-k}^k)$. This is denoted by $G_k\left(P_{Z_{-k}^k}, \Pi, f\right)$. In the next subsection, we explicitly derive this function.

The main idea behind our construction is to look at the worst case performance of a particular denoiser $f$ over all the channels in the uncertainty set $\Delta$. Since $G_k$ gives the performance of $f$ for a given channel $\Pi$, we can take the maximum over all the channels in $\Delta$. Define

$$J_k\left(P_{Z_{-k}^k}, \Delta, f\right) = \max_{\Pi \in \Delta} G_k\left(P_{Z_{-k}^k}, \Pi, f\right).$$

By definition, $J_k$ is the worst case performance of denoiser $f$ over all the channels in $\Delta$. Let $\mathcal{F}_k$ denote the set of all $k$-order sliding window denoisers. We now define the min-max denoiser,

$$f_{\text{MM}_k}[P_{Z_{-k}^k}, \Delta] = \arg\min_{f \in \mathcal{F}_k} J_k\left(P_{Z_{-k}^k}, \Delta, f\right). \tag{1}$$

By construction, $f_{\text{MM}_k}$ minimizes the worst expected loss over all channels in $\Delta$. Unfortunately, employment of this scheme requires knowledge of the noise-corrupted source distribution $P_{Z_{-k}^k}$, which is not given in this setting. Our approach is to employ $f_{\text{MM}_k}$ using an estimate of $P_{Z_{-k}^k}$. In particular, letting $\hat{Q}^{2k+1}[z^n]$ denote the $(2k+1)$-order empirical distribution induced by $z^n$, we look at the $n$-block denoiser defined by $f_{\text{MM}_k}[\hat{Q}^{2k+1}[z^n], \Delta]$.

Up to now in the development of our denoiser, the uncertainty set $\Delta$ remained unchanged. However, it is reasonable to assume that knowledge gained from our observations of the output processes **Z** can be used to modify the uncertainty set. In order to make this intuition more rigorous, we make use of the following definition. Given an observed output distribution $P_{Z^k}$, a channel $\Pi$ is said to be *k-feasible* if there exists a valid $k$-order distribution $P_{X^k}$ such that $\Pi * P_{X^k} = P_{Z^k}$.[7] As an example, we can look at a Bernoulli($p$) source corrupted by a binary symmetric channel with unknown crossover probability $\delta$, and assume $p$, $\delta < 1/2$. In this case, the output process will also be a Bernoulli source with parameter $q = p(1-\delta) + (1-p)\delta$. Then it is clear that for any $k$, no binary symmetric channel with crossover probability greater than $q$ is $k$-feasible. Similarly, all binary symmetric channels with crossover probability less than $q$ are $k$-feasible for all $k$. We shall say that a channel $\Pi$ is *feasible* with respect to the noise-corrupted source distribution $P_\mathbf{Z}$ if $\Pi$ is $k$-feasible with respect to $P_{Z^k}$ for all $k$.

---

[5]Similarly, we will use $\mathcal{S}^k(\mathcal{A})$ to denote the simplex on $k$-tuples on the alphabet $\mathcal{A}$. Also, $\mathcal{S}^{\infty}(\mathcal{A})$ will denote the set of all distribution on doubly infinite sequences that take value in $\mathcal{A}$. If no alphabet is given, the alphabet $\mathcal{A}$ is assumed.

[6]Throughout, given a random variable $X$, $P_X$ will be used to represent the associated probability law. Similar notation will also be used for vectors of random variables, such as $P_{X_{-k}^k}$ to denote the probability law associated with the vector $X_{-k}^k$. This will hold even for doubly infinite vectors like **Z**.

Using this concept of feasibility, given $P_{Z^k_{-k}}$, define

$$\mathcal{C}_k\left(P_{Z^k_{-k}}\right) = \left\{\Pi \in \mathcal{C}(\mathcal{A}) : \exists P_{X^k_{-k}} \in \mathcal{S}^{2k+1} \text{ s.t. } \Pi * P_{X^k_{-k}} = P_{Z^k_{-k}}\right\}, \quad (2)$$

where $\mathcal{C}(\mathcal{A})$ is the set of all invertible channels whose input and output take values in the alphabet $\mathcal{A}$. Recall that $\mathcal{S}^{2k+1}$ denotes the probability simplex on $(2k+1)$-tuples in $\mathcal{A}$. Therefore, $\mathcal{C}_k(P_{Z^k_{-k}})$ is simply the set of all $(2k+1)$-feasible channels with respect to the output distribution $P_{Z^k_{-k}}$. With a slight abuse of notation, we will also use $\mathcal{C}_k(P_\mathbf{Z})$ to represent $\mathcal{C}_k(P_{Z^k_{-k}})$. Furthermore, we will use $\mathcal{C}_\infty(P_\mathbf{Z})$ to denote the set of feasible channels, i.e. those channels which are in $\mathcal{C}_k(P_\mathbf{Z})$ for all $k$.

With our Bernoulli example in mind, we see that it need not be the case that given $P_{Z^k_{-k}}$, all the channels in $\Delta$ are $(2k+1)$-feasible. Hence we can rule out all channels in our uncertainty set $\Delta$ which are found not to be $(2k+1)$-feasible with respect to the observed output distribution. In other words, we can trim the uncertainty set down from $\Delta$ to $\Delta \cap \mathcal{C}_k(P_{Z^k_{-k}})$. This added information motivates the construction of our denoiser: We now define the $n$-block denoiser using the function $f_{\mathbb{MM}_k}$ from (1) by letting its estimate of $X_i$ be

$$\hat{X}_i \sim f_{\mathbb{MM}_k}\left[\hat{Q}^{2k+1}[z^n], \Delta \cap \mathcal{C}_l\left(\hat{Q}^{2l+1}[z^n]\right)\right](z^{i+k}_{i-k}). \quad (3)$$

Note that this denoiser depends on parameters $k$, $l$ and the a-priori uncertainty set $\Delta$. We denote this $n$-block denoiser by $\hat{X}^{n,k,l}_\Delta$. For the special case where we know $\Delta \subseteq \mathcal{C}_\infty(P_\mathbf{Z})$, let $\hat{X}^{n,k}_\Delta$ denote the denoiser defined by

$$\hat{X}_i \sim f_{\mathbb{MM}_k}\left[\hat{Q}^{2k+1}[z^n], \Delta\right](z^{i+k}_{i-k}). \quad (4)$$

## B  Construction of Denoiser

We now give a more detailed account of the construction of $\hat{X}^{n,k,l}_\Delta$ and $\hat{X}^{n,k}_\Delta$, and elaborate on technical details that arise in their derivation. Assume we are given a channel $\Pi \in \Delta$, a $(2k+1)$-order output distribution $P_{Z^k_{-k}}$, and a sliding window denoiser $f$. For a fixed two-sided context $Z^{-1}_{-k} = z^{-1}_{-k}$ and $Z^k_1 = z^k_1$, $P_{Z^k_{-k}}$ induces a conditional distribution on $Z_0$, denoted by $P_{Z_0|Z^{-1}_{-k}=z^{-1}_{-k}, Z^k_1=z^k_1}$ or, in short, $P_{Z_0|z^{-1}_{-k}, z^k_1}$.

We now wish to derive a function $F_k(P_{Z_0|z^{-1}_{-k}, z^k_1}, \Pi, f)$ which gives the expected loss, with respect to $\Lambda$, incurred when we estimate $X_0$ with the denoiser $f(Z^k_{-k})$ given that $Z^{-1}_{-k} = z^{-1}_{-k}$ and $Z^k_{-1} = z^k_{-1}$. Note that when $P_{Z^k_{-k}}$ is a channel output distribution and there exists an input distribution $P_{X^k_{-k}}$ such that

---
[7]Throughout, given a distribution on $k$-tuples $P_{X^k}$, $\Pi * P_{X^k}$ will denote the $k$-tuple distribution of the output of a DMC whose transition matrix is $\Pi$ and input has the $k$-order distribution $P_{X^k}$.

$\Pi * P_{X^k_{-k}} = P_{Z^k_{-k}}$, it is easy to show that $P_{Z_0|z^{-1}_{-k},z^k_1} = \Pi * P_{X_0|z^{-1}_{-k},z^k_1}$ (cf., e.g., [15, Section 3]). Therefore, the expected loss calculated by the function $F_k$ can be viewed as a twofold expectation, with respect to $P_{X_0|z^{-1}_{-k},z^k_1}$, and the denoiser. We can therefore write out $F_k$ as:

$$F_k\left(P_{Z_0|z^{-1}_{-k},z^k_1}, \Pi, f\right) = \sum_{x \in \mathcal{A}, z \in \mathcal{A}} \left[\Pi^{-T} P_{Z_0|z^{-1}_{-k},z^k_1}\right]_x \Pi(x, z) \left[\sum_{a \in \mathcal{A}} \Lambda(x, a) f\left([z^{-1}_{-k}, a, z^k_1]\right)[z]\right]$$

$$= \sum_{z \in \mathcal{A}} \sum_{x \in \mathcal{A}} \left[\Pi^{-T} P_{Z_0|z^{-1}_{-k},z^k_1}\right]_x \Pi(x, z) \left[\Lambda \cdot f\left([z^{-1}_{-k}, \cdot, z^k_1]\right)[z]\right]_x \quad (5)$$

$$= \sum_{z \in \mathcal{A}} \mathbf{1}^T \left[\Pi^{-T} P_{Z_0|z^{-1}_{-k},z^k_1} \odot \boldsymbol{\pi}_z \odot \left[\Lambda \cdot f\left([z^{-1}_{-k}, \cdot, z^k_1]\right)[z]\right]\right] \quad (6)$$

where:

- Given a channel $\Pi$ and $x, z \in \mathcal{A}$, $\Pi(x, z)$ denotes the probability the channel output is $z$ given the input is $x$. With a slight abuse of notation, $\Pi$ without an argument will denote the channel transition matrix. Similarly, $\Lambda$ without an argument will be used to denote the $|\mathcal{A}| \times |\mathcal{A}|$ matrix whose $(x, z)$-th entry is given by $\Lambda(x, z)$.

- $\left[\Pi^{-T} P_{Z_0|z^{-1}_{-k}}\right]_x$ denotes the $x$-th component of the column vector $\Pi^{-T} P_{Z_0|z^{-1}_{-k}}$.

- $f([z^{-1}_{-k}, a, z^k_1])[z]$ is the $z$-th element of the $|\mathcal{A}|$-dimensional simplex member $f([z^{-1}_{-k}, a, z^k_1])$, and $f([z^{-1}_{-k}, \cdot, z^k_1])[z]$ is the column vector whose $a$-th component is $f([z^{-1}_{-k}, a, z^k_1])[z]$. Recall that a denoiser $f$ is a mapping $\mathcal{A}^{2k+1} \mapsto \mathcal{S}$.

- $\mathbf{1}$ denotes the "all ones" $|\mathcal{A}|$-dimensional column vector.

- $\odot$ Denotes the Hadamard product, that is the component-wise multiplication.

- $\boldsymbol{\pi}_z$ Denotes the $|\mathcal{A}|$-dimensional column vector whose $a$-th component is $\Pi(a, z)$.

Equipped with the function $F_k$, we can now construct $G_k$. Recall for a given channel $\Pi$, a $(2k+1)$-order output distribution $P_{Z^k_{-k}}$, and a denoiser $f$, $G_k$ calculates the expected loss with respect to $\Lambda$. Hence $F_k$ can be thought of as $G_k$ conditioned on a particular context $z^{-1}_{-k}$ and $z^k_1$. It follows that

$$G_k(P_{Z^k_{-k}}, \Pi, f) = \sum_{z^{-1}_{-k}, z^k_1 \in \mathcal{A}^k} F_k\left(P_{Z_0|z^{-1}_{-k},z^k_1}, \Pi, f\right) P_{Z^k_{-k}}\left\{Z^{-1}_{-k} = z^{-1}_{-k}, Z^k_1 = z^k_1\right\}, \quad (7)$$

where $P_{Z^k_{-k}}\left\{Z^{-1}_{-k} = z^{-1}_{-k}, Z^k_1 = z^k_1\right\}$ is the probability under the law $P_{Z^k_{-k}}$ that $Z^{-1}_{-k} = z^{-1}_{-k}$ and $Z^k_1 = z^k_1$. Substituting (6) in (7) and simplifying gives

$$G_k(P_{Z^k_{-k}}, \Pi, f) = \sum_{z^{-1}_{-k}, z^k_1 \in \mathcal{A}^k} \sum_{z \in \mathcal{A}} \mathbf{1}^T \left[\Pi^{-T} P_{Z_0|z^{-1}_{-k},z^k_1} \odot \boldsymbol{\pi}_z \odot \left[\Lambda \cdot f([z^{-1}_{-k}, \cdot, z^k_1])[z]\right]\right] P_{Z^k_{-k}}\left\{Z^{-1}_{-k} = z^{-1}_{-k}, Z^k_1 = z^k_1\right\}. \quad (8)$$

Following the development in Section 3A, we now use $G_k$ in the construction of $J_k$:

$$J_k\left(P_{Z^k_{-k}}, \Delta, f\right) = \max_{\Pi \in \Delta} G_k\left(P_{Z^k_{-k}}, \Pi, f\right).$$

We make the following two observations. The function $J_k(P_{Z_{-k}^k}, \Delta, f)$ is continuous in $f$, i.e. continuous in the space of all $(2k+1)$-order sliding window denoisers. This is an easily verified consequence of the definition of $G_k$. The second observation requires the construction of a metric, $\rho$, between sets of channels. Recall that $\mathcal{C}(\mathcal{A})$ denotes the set of invertible channels whose input and output take values in the alphabet $\mathcal{A}$. For nonempty $A, B \subseteq \mathcal{C}(\mathcal{A})$ we define

$$\rho(A, B) = \sup_{a \in A} \inf_{b \in B} ||a - b|| + \sup_{b \in B} \inf_{a \in A} ||a - b||,$$

where $||\cdot||$ denotes the $L^\infty$ norm. With respect to the metric $\rho$, $J_k(P_{Z_{-k}^k}, \Delta, f)$ is uniformly continuous in $\Delta$. More specifically, for all $\Delta' \subset \Delta$,

$$\left| J_k\left(P_{Z_{-k}^k}, \Delta, f\right) - J_k\left(P_{Z_{-k}^k}, \Delta', f\right) \right| \leq \phi_k(\rho(\Delta, \Delta')), \tag{9}$$

for some $\phi_k$, independent of $P_{Z_{-k}^k}$ and $f$, such that $\phi_k(\varepsilon) \downarrow 0$ as $\varepsilon \downarrow 0$. For example,

$$\phi_k(\varepsilon) = \left[|\mathcal{A}|^{2k+1} \Lambda_{\max} \max_{\Pi \in \Delta} ||\Pi^{-1}||\right] \varepsilon \tag{10}$$

is readily verified to satisfy (9).

Continuing the development, as per our previous definition,

$$f_{\mathbb{MM}_k}[P_{Z_{-k}^k}, \Delta] = \arg \min_{f \in \mathcal{F}_k} J_k\left(P_{Z_{-k}^k}, \Delta, f\right)$$

selecting an arbitrary achiever when it is not unique. Note that the minimum is achieved since, as observed, $J_k$ is continuous in $f$ and the space of all $(2k+1)$-order sliding window denoisers is compact. Equation (3) and (4) then complete the construction of the denoisers.

## C  Binary Alphabet

Before moving on, it may be illustrative to explore the form of $\hat{X}_\Delta^{n,k,l}$ for the binary case. In particular, we will look at the case of denoising a binary signal corrupted by an unknown Binary Symmetric Channel (BSC) with respect to the Hamming loss. We suppose it is known that the BSC lies in some finite set $\Delta$. We will assume that all the channels in $\Delta$ have a crossover probability less than $1/2$.

The first step in constructing our binary denoiser is finding the binary version of $F_k$. Let us fix a particular context $z_{-k}^{-1}$ and $z_1^k$. As we recall from (6), $F_k$ is a function of a distribution $P_{Z_0|z_{-k}^{-1}, z_1^k}$, a channel $\Pi$, and a denoiser $f$. In the binary case, $P_{Z_0|z_{-k}^{-1}, z_1^k}$ is completely specified by the conditional probability that $Z_0 = 1$. We will denote this probability as $\alpha(z_{-k}^{-1}, z_1^k)$. The channel is a BSC and therefore defined by its crossover probability, denoted by $\delta < 1/2$. Also recall that a denoiser $f$ is a mapping from $\{0,1\} \mapsto \mathcal{S}(\{0,1\})$. Hence for our two-sided context, $f$ can be completely defined by the probability assigned to 1 given $Z_0 = 1$, denoted by $d_1(z_{-k}^{-1}, z_1^k)$, and the probability assigned to 1 given $Z_0 = 0$, denoted by $d_0(z_{-k}^{-1}, z_1^k)$. Finally, recall that $F_k$ measures the expected loss, here with respect to the Hamming loss, incurred when we estimate $X_0$ with $f(Z_{-k}^k)$ given that $Z_{-k}^{-1} = z_{-k}^{-1}$ and $Z_1^k = z_1^k$. With this in mind, we write out $F_k$ for the binary case as

$$\begin{aligned}
F_k(P_{Z_0|z_{-k}^{-1},z_1^k}, \Pi, f) &= F_k(\alpha, \delta, [d_0, d_1]) \\
&= \Lambda(0,0)\left[\Pr\{X_0=0, Z_0=1\}\bar{d}_1 + \Pr\{X_0=0, Z_0=0\}\bar{d}_0\right] \\
&\quad + \Lambda(0,1)\left[\Pr\{X_0=0, Z_0=1\}d_1 + \Pr\{X_0=0, Z_0=0\}d_0\right] \\
&\quad + \Lambda(1,0)\left[\Pr\{X_0=1, Z_0=1\}\bar{d}_1 + \Pr\{X_0=1, Z_0=0\}\bar{d}_0\right] \\
&\quad + \Lambda(1,1)\left[\Pr\{X_0=1, Z_0=1\}d_1 + \Pr\{X_0=1, Z_0=1\}d_0\right] \\
&= \left[\Pr\{X_0=0, Z_0=1\}d_1 + \Pr\{X_0=0, Z_0=0\}d_0\right] \\
&\quad + \left[\Pr\{X_0=1, Z_0=1\}\bar{d}_1 + \Pr\{X_0=1, Z_0=0\}\bar{d}_0\right] \\
&= \left[\frac{\delta(1-\alpha-\delta)}{1-2\delta}d_1 + \frac{\bar{\delta}(1-\alpha-\delta)}{1-2\delta}d_0\right] \\
&\quad + \left[\frac{\bar{\delta}(\alpha-\delta)}{1-2\delta}\bar{d}_1 + \frac{\delta(\alpha-\delta)}{1-2\delta}\bar{d}_0\right] \\
&= \frac{\delta(1-\alpha-\delta)d_1 + \bar{\delta}(1-\alpha-\delta)d_0 + \bar{\delta}(\alpha-\delta)\bar{d}_1 + \delta(\alpha-\delta)\bar{d}_0}{1-2\delta}, \quad (11)
\end{aligned}$$

where we dropped $\alpha(z_{-k}^{-1}, z_1^k)$, $d_1(z_{-k}^{-1}, z_1^k)$, and $d_0(z_{-k}^{-1}, z_1^k)$ dependence of $z_{-k}^{-1}$ and $z_1^k$ for notational compactness. Using (11), we can then follow the construction in Section 3B to derive the binary version of the denoiser $\hat{X}_\Delta^{n,k,l}$. The practical implementation of this denoiser is discussed in detail in [8].

## 4 Performance Criterion

In the setting of [15], the known channel setting, performance is measured by expected loss and optimal performance is characterized via the Bayes Envelope. In that setting, with the expected loss performance measure, a denoiser which achieves the Bayes Envelope is optimal. However, as the following example illustrates, this performance measure and guarantee are not relevant for the unknown channel setting.

**Example 1** *Let $\mathbf{Z}$ be a binary source, $\mathbf{X}$, corrupted by a BSC with unknown crossover probability $\delta \in \Delta = \{.1, .2\}$. Furthermore, $\mathbf{Z}$ is known to be a Bernoulli process with parameter $1/4$. Therefore, we know that $\mathbf{X}$ is also a Bernoulli process with parameter $\alpha < 1/4$. We want to reconstruct $X^n$ from $Z^n$ with respect to the Hamming loss function. Let us examine the two possible cases:*

1. *The channel crossover probability $\delta$, is .1. Since the Bernoulli process $\mathbf{Z}$ has parameter $1/4$, we determine that $\alpha = .1875$. Since $\alpha > \delta$, it is readily seen that in order to minimize loss, we should reconstruct $X_i$ with the observation $Z_i$. This scheme achieves the Bayes Envelope for a BSC with $\delta = .1$.*

2. *The channel crossover probability $\delta$, is .2. Since the Bernoulli process $\mathbf{Z}$ has parameter $1/4$, we determine that $\alpha = .0833$. Since $\alpha < \delta$, it is readily seen that in order to minimize expected loss, we should reconstruct $X_i$ with $0$ regardless of the observed $Z_i$. The optimality of this reconstruction scheme stems from the fact that when $\alpha < \delta$, an observed $1$ in the channel output is more likely to be caused by the*

*BSC than the source. Similarly to our previous case, this scheme achieves the Bayes Envelope for a BSC with $\delta = .2$.*

*We also observe that the optimal scheme for one case is suboptimal for the other.*

From Example 1, we see that although one can achieve the Bayes Envelope for each channel in the uncertainty set, there may not be one denoiser that can achieve the Bayes Envelope for each channel simultaneously. In particular, there does not exist a denoiser which is simultaneously optimal for the two possible channels in Example 1. It is therefore problematic to compare various denoisers in the unknown channel setting using expected loss as a performance measure. How would one rank the two denoising schemes suggested in Example 1? Each scheme is optimal for one of the two possible channels, but suboptimal for the other. This difficulty also leads to an ambiguity in defining an optimal denoiser.

Clearly, a new performance measure is needed for our setting of the unknown channel. Without any prior on the uncertainty set, a natural performance measure which is applicable in this setting is a min-max, or worst case measure. In other words, we look at the worst case expected loss of a denoiser across all possible channels in the uncertainty set $\Delta$. Such a performance measure would take into account the entire uncertainty set. With this is mind, we can define our performance measure. Before doing so we need to introduce some notation. For $x^n, z^n \in \mathcal{A}^n$, given a $k$-order sliding window denoiser $f$ we denote

$$L_f(x^n, z^n) = \frac{1}{n} \sum_{t=k+1}^{n-k} \sum_{a \in \mathcal{A}} \Lambda(x_t, a) f\left(z_{t-k}^{t+k}\right)[a], \tag{12}$$

the normalized loss[8] when employing the sliding window denoiser $f$. Here we make the assumption that $k < n$. Furthermore, given a channel $\Pi$ and a source distribution $P_{\mathbf{X}}$, $P_{[P_{\mathbf{X}},\Pi]}$ will denote the joint distribution on $(\mathbf{X}, \mathbf{Z})$ when $\mathbf{X} \sim P_{\mathbf{X}}$ and $\mathbf{Z}$ is the output of the channel $\Pi$ with input $\mathbf{X}$. Given an uncertainty set $\Delta$, we now define our performance measure as follows:

$$\mathcal{L}_f^{(n)}(P_{\mathbf{Z}}, \Delta, \mathbf{Z}) = \sup_{\{(P_{\mathbf{X}},\Pi):\Pi \in \Delta, \Pi * P_{\mathbf{X}} = P_{\mathbf{Z}}\}} E_{[P_{\mathbf{X}},\Pi]}\left[L_f(X^n, Z^n)|\mathbf{Z}\right], \tag{13}$$

where $E_{[P_{\mathbf{X}},\Pi]}[\ \cdot\ |\mathbf{Z}]$ denotes the conditional expectation, with respect to the joint distribution $P_{[P_{\mathbf{X}},\Pi]}$, given $\mathbf{Z}$. In words, for a given denoiser $f$, an uncertainty set $\Delta$, and the noise-corrupted source $\mathbf{Z}$, $\mathcal{L}_f(P_{\mathbf{Z}}, \Delta, \mathbf{Z})$ is the worst case expected loss of the denoiser $f$ over all feasible channels in the uncertainty set $\Delta$, given $\mathbf{Z}$. The performance measure in (13) is conditioned on the noise-corrupted sequence $\mathbf{Z}$ since it seems natural that the performance of a denoiser be determined on the basis of the actual source realization, rather than merely on its distribution. Although the performance measure is defined using this conditioning, in Sections 5 and 6, performance guarantees are given for both the conditional performance measure and a non-conditional version.

Equipped with our new performance measure, we can now compare the two denoising schemes suggested in Example 1. Let $f_1$ and $f_2$ denote the denoising scheme of Case 1 and Case 2, respectively, i.e. $f_1$ is the "say what you see" scheme and $f_2$ is the "say all zeros" scheme. Furthermore, given the Bernoulli process **Z**, let $N_1(Z^n)$ be the frequency of ones in $Z^n$. We see that, for any $n$,

$$\begin{aligned}
\mathcal{L}_{f_1}^{(n)}(P_\mathbf{Z}, \{.1, .2\}, \mathbf{Z}) &= \max_{\delta \in \{.1, .2\}} E\left[\frac{1}{n}\sum_{i=1}^n \mathbb{1}_{X_i \neq Z_i} \Big| \mathbf{Z}\right] \\
&= \max_{\delta \in \{.1, .2\}} \frac{1}{n}\sum_{i=1}^n E\left[\mathbb{1}_{X_i \neq Z_i} | \mathbf{Z}\right] \\
&= \max_{\delta \in \{.1, .2\}} \frac{1}{n}\sum_{i=1}^n E\left[\mathbb{1}_{X_i \neq Z_i} | Z_i\right] \\
&= \max_{\delta \in \{.1, .2\}} N_1(Z^n)\Pr\{X_0 \neq Z_0|Z_0 = 1\} + (1 - N_1(Z^n))\Pr\{X_0 \neq Z_0|Z_0 = 0\} \\
&= \max_{\delta \in \{.1, .2\}} N_1(Z^n)\frac{\delta \Pr\{X_0 = 0\}}{\Pr\{Z_i = 1\}} + (1 - N_1(Z^n))\frac{\delta \Pr\{X_0 = 1\}}{\Pr\{Z_i = 0\}} \\
&= \max_{\delta \in \{.1, .2\}} \left(\frac{N_1(Z^n)}{\Pr\{Z_i = 1\}}\Pr\{X_0 = 0\} + \frac{1 - N_1(Z^n)}{\Pr\{Z_i = 0\}}\Pr\{X_0 = 1\}\right)\delta
\end{aligned}$$

and that

$$\begin{aligned}
\mathcal{L}_{f_2}^{(n)}(P_\mathbf{Z}, \{.1, .2\}, \mathbf{Z}) &= \max_{\delta \in \{.1, .2\}} N_1(Z^n)\Pr\{X_0 = 1|Z_0 = 1\} + (1 - N_1(Z^n))\Pr\{X_0 = 1|Z_0 = 0\} \\
&= \max_{\delta \in \{.1, .2\}} \frac{N_1(Z^n)}{\Pr\{Z_0 = 1\}}(1 - \delta)\Pr\{X_0 = 1\} + \frac{(1 - N_1(Z^n))}{\Pr\{Z_0 = 0\}}\delta \Pr\{X_0 = 1\}.
\end{aligned}$$

---
[8]Up to the "edge-effects" associated with indices $t$ outside the range $k + 1 \leq t \leq n - k$ that will be asymptotically inconsequential in our analysis (which will assume $k \ll n$).

The strong law of large numbers states that as $n \to \infty$, $N_1(Z^n)$ converges to $\Pr\{Z_0 = 1\}$ w.p. 1. Therefore, for large $n$

$$\begin{aligned}
\mathcal{L}_{f_1}^{(n)}(P_\mathbf{Z}, \{.1, .2\}, \mathbf{Z}) &\approx .2 \\
\mathcal{L}_{f_2}^{(n)}(P_\mathbf{Z}, \{.1, .2\}, \mathbf{Z}) &\approx .1875
\end{aligned}$$

with high probability. Can we find a denoiser that does better than the two suggested in Example 1?

One possible way to improve denoiser performance in Example 1 is to time share between the two suggested denoisers schemes, "say what you see" and "say all zeros." For $\gamma \in [0, 1]$, let $f^{(\gamma)}$ be a denoiser which at each reconstruction implements "say what you see" with probability $\gamma$ and "say all zeros" with probability $1 - \gamma$. To simplify our calculations, we will assume that $n$ is large enough such that $N_1(Z^n)$ is close to $\Pr\{Z_0 = 1\}$ with high probability. We can now calculate the performance of this denoiser as follows:

$$\begin{aligned}
\mathcal{L}_{f^{(\gamma)}}^{(n)}(P_\mathbf{Z}, \{.1, .2\}, \mathbf{Z}) &\approx \max_{\delta \in \{.1, .2\}} \gamma \Pr\{X_i \neq Z_i\} + (1 - \gamma)\Pr\{X_i = 0\} \\
&= \max\{.1\gamma + .1875(1 - \gamma), .2\gamma + .0833(1 - \gamma)\} \\
&= \max\{.0875\gamma + .1875, .1168\gamma + .0833\},
\end{aligned}$$

with high probability. We can then find the best such denoiser by finding the $\gamma$ which minimizes the worst case loss. It is easily seen that, with high probability,

$$\begin{aligned}
\min_{\gamma \in [0,1]} \mathcal{L}_{f^{(\gamma)}}^{(n)}(P_\mathbf{Z}, \{.1, .2\}, \mathbf{Z}) &\approx \min_{\gamma \in [0,1]} \max\{.0875\gamma + .1875, .1168\gamma + .0833\} \\
&= .1428,
\end{aligned}$$

and that the minimum is achieved by $\gamma = .5101$.

We see then that, for typical[9] $\mathbf{z}$, $f^{(.5101)}$ is a better denoiser than $f_1$ and $f_2$, but what is the best denoiser? To answer this question, we develop the concept of an optimal denoiser under the worst case loss performance measure defined in (13). First, recall that $\mathcal{F}_k$ denotes the set of all $k$-order sliding window denoisers. Now define

$$\mu_k^{(n)}(P_{\mathbf{Z}}, \Delta, \mathbf{Z}) = \min_{f \in \mathcal{F}_k} \mathcal{L}_f^{(n)}(P_{\mathbf{Z}}, \Delta, \mathbf{Z}), \tag{14}$$

$$\mu_k(P_{\mathbf{Z}}, \Delta, \mathbf{Z}) = \limsup_{n \to \infty} \mu_k^{(n)}(P_{\mathbf{Z}}, \Delta, \mathbf{Z}). \tag{15}$$

In words, $\mu_k^{(n)}(P_{\mathbf{Z}}, \Delta, \mathbf{Z})$ is the performance of the best $k$-order sliding window denoiser operating on blocks of size $n$.[10] We then take $n \to \infty$ to define $\mu_k(P_{\mathbf{Z}}, \Delta, \mathbf{Z})$, the performance of the best $k$-order sliding window denoiser. Finally we let $k \to \infty$ and define the "sliding window minimum loss,"

$$\mu(P_{\mathbf{Z}}, \Delta, \mathbf{Z}) = \lim_{k \to \infty} \mu_k(P_{\mathbf{Z}}, \Delta, \mathbf{Z}), \tag{16}$$

where the limit is actually an infimum since for every $\mathbf{Z}$, $\mu_k(P_{\mathbf{Z}}, \Delta, \mathbf{Z})$ is point wise non-increasing with $k$. In words, $\mu(P_{\mathbf{Z}}, \Delta, \mathbf{Z})$ is the performance of the best sliding window denoiser of any order. Hence $\mu(P_{\mathbf{Z}}, \Delta, \mathbf{Z})$ is a bound on the performance of any sliding window denoiser. We denote a denoiser as optimal if it achieves this performance bound $P_{\mathbf{Z}}$-a.s., the need for an almost sure statement comes from the fact that both the performance bound and measure depend on the source realization. Surprisingly, it can be shown that the denoiser $f^{(.5101)}$ defined above is optimal for the Example 1, i.e., with high probability comes close to attaining the minimum in (14) for all $k$. This is due to the memorylessness of the source in Example 1.

One can consider $\mu(P_{\mathbf{Z}}, \Delta, \mathbf{Z})$ defined in (14) as a kind of analogue in our setup to the "sliding window minimum loss" of [15, Section 5] which, in turn, is analogous to the finite-state compressibility of [18], the finite-state predictability of [7], and the conditional finite-state predictability of [14].

## 5  Performance Guarantees

In this section we present a result on the performance of the algorithm presented in Section 3 with respect to the performance measure discussed in the previous section.

Throughout this section the uncertainty set $\Delta$ is assumed to be finite. Additionally, to isolate the main issue of minimizing the worst case performance from the issue of estimating the set of channels in the uncertainty set, we limit our first theorem to the case where all channels in the uncertainty set are known to be feasible, namely they satisfy $\Delta \subseteq \mathcal{C}_{\infty}(P_{\mathbf{Z}})$.

---

[9]In particular, all $\mathbf{z}$ with $\lim_{n \to \infty} N_1(z^n) = 1/4$.

[10]Although $\mu_k^{(n)}$ is defined as a minimum over an uncountable set, it is easily seen to be point-wise equal to $\min_{f: \mathcal{A}^{2k+1} \mapsto \mathcal{S}_{\mathbb{Q}}} \mathcal{L}_f^{(n)}(P_{\mathbf{Z}}, \Delta, \mathbf{Z})$, where we use $\mathcal{S}_{\mathbb{Q}}$ to denote the subset of $\mathcal{S}$ consisting of distributions with rational components. The latter is a minimum over a countable set of random variables and hence measurable.

**Theorem 1** *Let*
$$\hat{X}^n_{univ} = \hat{X}^{n,k_n}_{\Delta},$$
*where on the right side is the n-block denoiser defined in (4) and let $\{k_n\}$ be any sequence satisfying $k_n \leq \frac{\ln n}{16 \ln |\mathcal{A}|}$. For any output distribution $P_\mathbf{Z}$ such that $\Delta \subseteq \mathcal{C}_\infty(P_\mathbf{Z})$,*

$$\lim_{n \to \infty} \left[ \mathcal{L}_{\hat{X}^n_{univ}}(P_\mathbf{Z}, \Delta, \mathbf{Z}) - \mu^{(n)}_{k_n}(P_\mathbf{Z}, \Delta, \mathbf{Z}) \right] = 0 \qquad P_\mathbf{Z} - a.s. \tag{17}$$

We defer the proof of Theorem 1 to the appendix.

*Remarks:* Note that beyond the stipulation $\Delta \subseteq \mathcal{C}_\infty(P_\mathbf{Z})$, no other assumption is made on $P_\mathbf{Z}$, not even stationarity. Note also that, as a direct consequence of (14), we have for each $n$ and all possible realizations of $\mathbf{Z}$,

$$\mathcal{L}_{\hat{X}^n_{univ}}(P_\mathbf{Z}, \Delta, \mathbf{Z}) \geq \mu^{(n)}_{k_n}(P_\mathbf{Z}, \Delta, \mathbf{Z}).$$

Thus, the non-trivial part of (17) is that

$$\limsup_{n \to \infty} \left( \mathcal{L}_{\hat{X}^n_{univ}}(P_\mathbf{Z}, \Delta, \mathbf{Z}) - \mu^{(n)}_{k_n}(P_\mathbf{Z}, \Delta, \mathbf{Z}) \right) \leq 0 \qquad P_\mathbf{Z} - a.s.$$

An immediate consequence of Theorem 1 is:

**Corollary 1** *Let the setting of Theorem 1 hold and $k_n \to \infty$. For any $P_\mathbf{Z}$ such that $\Delta \subseteq \mathcal{C}_\infty(P_\mathbf{Z})$*

$$\limsup_{n \to \infty} \mathcal{L}_{\hat{X}^n_{univ}}(P_\mathbf{Z}, \Delta, \mathbf{Z}) \leq \mu(P_\mathbf{Z}, \Delta, \mathbf{Z}) \qquad P_\mathbf{Z} - a.s. \tag{18}$$

*Proof:* We have $P_\mathbf{Z}$-a.s.,

$$\limsup_{n \to \infty} \mathcal{L}_{\hat{X}^n_{univ}}(P_\mathbf{Z}, \Delta, \mathbf{Z}) = \limsup_{n \to \infty} \mu^{(n)}_{k_n}(P_\mathbf{Z}, \Delta, \mathbf{Z})$$
$$\leq \mu(P_\mathbf{Z}, \Delta, \mathbf{Z}), \tag{19}$$

where the equality follows from Theorem 1. The inequality comes from the fact that for any fixed $k$, since $k_n$ increases without bound,

$$\limsup_{n \to \infty} \mu^{(n)}_{k_n}(P_\mathbf{Z}, \Delta, \mathbf{Z}) \leq \mu_k(P_\mathbf{Z}, \Delta, \mathbf{Z}).$$

Therefore the left side is also upper bounded by $\inf_{k \geq 1} \mu_k(P_\mathbf{Z}, \Delta, \mathbf{Z}) = \mu(P_\mathbf{Z}, \Delta, \mathbf{Z})$.

□

Corollary 1 states that asymptotically, in $n$ and the window size, the sliding window denoiser of Section 3 achieves the performance bound $\mu(P_\mathbf{Z}, \Delta, \mathbf{Z})$ $P_\mathbf{Z}$-a.s. The denoising scheme is therefore asymptotically optimal with respect to the worst case performance measure described in Section 4.

We also establish the following consequence of Theorem 1.

**Corollary 2** *Let $P_\mathbf{Z}$ be stationary and ergodic, $\Delta$ be finite, and $\hat{X}^n_{univ}$ be defined as in Theorem 1 with $k_n \equiv k$. If $\Delta \subseteq \mathcal{C}_\infty(P_\mathbf{Z})$, then*

$$\lim_{n \to \infty} \left| \max_{\{(P_\mathbf{X}, \Pi): \Pi \in \Delta, \Pi * P_\mathbf{X} = P_\mathbf{Z}\}} E_{[P_\mathbf{X}, \Pi]} \left[ L_{\hat{X}^n_{univ}}(X^n, Z^n) \right] - \min_{f \in \mathcal{F}_k} \max_{\{(P_\mathbf{X}, \Pi): \Pi \in \Delta, \Pi * P_\mathbf{X} = P_\mathbf{Z}\}} E_{[P_\mathbf{X}, \Pi]} \left[ L_f(X^n, Z^n) \right] \right| = 0.$$

For proof of Corollary 2, see the appendix.

Note that the difference between the kind of statement in Theorem 1 and that in Corollary 2 is that in the latter we omit the conditioning on the noise-corrupted sequence **Z**. The latter can be viewed as the analogue of our setting to the expectation results of [15], while the statement of Theorem 1 is more in the spirit of the semi-stochastic setting of [15].

## 6 Performance Guarantees For the General Case

In Section 5, we assumed that $|\Delta|$ was finite and that all channels in $\Delta$ are feasible. These two assumptions allowed us to avoid a few technicalities. In this section, we will remove these assumptions and extend the performance guarantees of Section 5 to the case where $\Delta$ is an infinite set, and we no longer require that $\Delta \subseteq \mathcal{C}_\infty(P_\mathbf{Z})$. To preserve the concept of invertibility, we require that $\max_{\Pi \in \Delta} ||\Pi^{-1}||$ be finite.

Before continuing, it is important to identify the issues that arise when we remove these two key assumptions. In (13), our performance measure $\mathcal{L}$ is defined to be the supremum of $E_{[P_\mathbf{X}, \Pi]}[L_f(X^n, Z^n)|\mathbf{Z}]$ over the set of feasible channels in $\Delta$. Although $E_{[P_\mathbf{X}, \Pi]}[L_f(X^n, Z^n)|\mathbf{Z}]$ is a measurable function for each $\Pi \in \Delta$, if $\Delta$ is an uncountable set, we are no longer assured that the supremum in (13) is measurable. Initially, to avoid this complication we made the assumption of $|\Delta|$ being finite.

To deal with this measurability issue in the development of $\mathcal{L}$, one may consider those channels in $\Delta$ which have rational transition matrices. Let $\mathcal{Q}(\mathcal{A})$ be the subset of channels in $\mathcal{C}(\mathcal{A})$ whose transition matrices have rational components. Then given an uncountable uncertainty set $\Delta$, we can look at $\mathcal{L}_f^{(n)}(P_\mathbf{Z}, \Delta \cap \mathcal{Q}(\mathcal{A}), \mathbf{Z})$. Since $\Delta \cap \mathcal{Q}(\mathcal{A})$ is a countable set, we are assured that $\mathcal{L}_f^{(n)}(P_\mathbf{Z}, \Delta \cap \mathcal{Q}(\mathcal{A}), \mathbf{Z})$ is well defined. Using this modification, we can extend the definition of $\mu_k$ and $\mu$. Similarly, we can use this approach in the construction of our denoiser $\hat{X}_\Delta^{n,k,l}$. We therefore assume that $\Delta \subseteq \mathcal{Q}(\mathcal{A})$.

The other assumption made in Section 5 is that all channels in $\Delta$ are feasible. We can remove this condition if $\Delta$ is sufficiently well behaved in the following sense:

**Assumption 1** *Given a set $A$, let $A^-$ denote its closure. For every stationary process $\mathbf{U}$,*

$$\bigcap_{l=1}^\infty \left(\Delta \cap \mathcal{C}_l(P_{U_{-l}^l})\right)^- = \left(\bigcap_{l=1}^\infty \Delta \cap \mathcal{C}_l(P_{U_{-l}^l})\right)^-$$

*and*

$$\rho\left(\Delta \cap \mathcal{C}_\infty(\mathbf{U}), \Delta \cap \mathcal{C}_l(\mathbf{U})\right)$$

*is continuous in $\mathbf{U}$ for all $l$.*

**Assumption 2** *For each $l$ there exists a function $b_l$ satisfying $b_l(\varepsilon) \downarrow 0$ as $\varepsilon \downarrow 0$ and*

$$\rho\left(\Delta \cap \mathcal{C}_l(P_{U_{-l}^l}), \Delta \cap \mathcal{C}_l(P'_{U_{-l}^l})\right) \leq b_l(\|P_{U_{-l}^l} - P'_{U_{-l}^l}\|). \tag{20}$$

Assumption 1 imposes a structural constraint on $\Delta$ while Assumption 2 gives us a form of continuity. To illustrate these two assumptions, let us explore the binary case. Let $\Delta$ consist of all BSCs with rational crossover probability less than some $\delta_0 < 1/2$. It is easy to see that any such $\Delta$ satisfies Assumption 1. Furthermore, Assumption 2 is satisfied with $b_l(\varepsilon) = \frac{\varepsilon}{(1-2\delta_0)^{2l}}$. More generally, if $\Delta$ consists of all channels in $\mathcal{Q}(\mathcal{A})$ within a certain radius of the noise free channel, then

$$b_l(\varepsilon) = \varepsilon(\max_{\Pi \in \Delta}||\Pi^{-1}||)^{|\mathcal{A}|^l} \tag{21}$$

satisfies Assumption 2.

Before we state the next performance guarantee, we need to introduce the notion of $\psi$-mixing. Roughly, the $i$-th $\psi$-mixing coefficient of a stationary source $P_{\mathbf{Z}}$ is defined as the maximum value of the distance between the value 1 and the Radon–Nikodym derivative between $P_{Z^0_{-\infty},Z^\infty_i}$ and the product distribution $P_{Z^0_{-\infty}} \times P_{Z^\infty_i}$ (cf. [3] for a rigorous definition). In our finite-alphabet setting, the $i$-th $\psi$-mixing coefficient associated with a given stationary source $P_{\mathbf{Z}}$ is more simply given by

$$\sup_{k,j>0} \max_{\left\{z^0_{-k},z^j_i : P_{Z^0_{-k}}(z^0_{-k})P_{Z^j_i}(z^j_i) \neq 0\right\}} \left| \frac{P_{Z^0_{-k},Z^j_i}(z^0_{-k},z^j_i)}{P_{Z^0_{-k}}(z^0_{-k})P_{Z^j_i}(z^j_i)} - 1 \right|.$$

Qualitatively, the $\psi$-mixing coefficients are a measure of the effective memory of a process. For a given sequence of nonnegative reals $\{\psi_i\}$ we let $\tilde{\mathcal{S}}_{\{\psi_i\}}$ denote all stationary sources whose $i$-th $\psi$-mixing coefficient is bounded above by $\psi_i$ for all $i$.

**Theorem 2** *Let $\{\psi_i\}$ be a sequence of nonnegative reals with $\psi_i \to 0$ and let $\Delta \subseteq \mathcal{Q}(\mathcal{A})$ satisfy Assumptions 1 and 2. There exists an unbounded sequences $\{l_n\}$ and $\{k_n\}$ such that if*

$$\hat{X}^n_{univ} = \hat{X}^{n,k_n,l_n}_\Delta,$$

*then for any $P_{\mathbf{Z}} \in \tilde{\mathcal{S}}_{\{\psi_i\}}$ and any sequence $\{\Delta_n\}$ with $\Delta_n \subseteq \Delta$ and $|\Delta_n| = O\left(e^{\sqrt{n}}\right)$*

$$\limsup_{n \to \infty} \left[ \mathcal{L}_{\hat{X}^n_{univ}}(P_{\mathbf{Z}}, \Delta_n, \mathbf{Z}) - \mu^{(n)}_{k_n}(P_{\mathbf{Z}}, \Delta, \mathbf{Z}) \right] \leq 0 \qquad P_{\mathbf{Z}} - a.s. \tag{22}$$

The proof of Theorem 2 makes use of a more general result, Lemma 7. Lemma 7 and the proof of Theorem 2 can be found in the appendix.

*Remarks:*

- The explicit dependence of $\{l_n\}$ and $\{k_n\}$ on $\{\psi_i\}$ is given in the proof.

- If $\psi_i = e^{-i\rho}$ for some $\rho > 0$ then any $w_n = o(\log n)$ will do.

- Any Markov source of any order with no restricted transitions, as well as any finite-state hidden Markov process whose underlying state sequence has no restricted transitions is exponentially mixing, i.e., belongs to $\tilde{\mathcal{S}}_{\{\psi_i\}}$ with $\psi_i = e^{-i\rho}$ for some $\rho > 0$ (cf. [6]).

Analogously as was done in Corollary 2, we can extend the results of Theorem 2 as follows:

**Proposition 1** *Let $\{\psi_i\}$ be a sequence of nonnegative reals with $\psi_i \to 0$ and assume finite $\Delta$. There exists unbounded sequences $\{l_n\}$ and $\{k_n\}$ such that if*

$$\hat{X}^n_{univ} = \hat{X}^{n,k_n,l_n}_\Delta,$$

*then for any $P_\mathbf{Z} \in \tilde{\mathcal{S}}_{\{\psi_i\}}$*

$$\lim_{n\to\infty} \left| \max_{\{(P_\mathbf{X},\Pi):\Pi\in\Delta, P_\mathbf{X}*\Pi=P_\mathbf{Z}\}} E_{[P_\mathbf{X},\Pi]}\left[L_{\hat{X}^n_{univ}}(X^n,Z^n)\right] - \min_{f\in\mathcal{F}_{k_n}} \max_{\{(P_\mathbf{X},\Pi):\Pi\in\Delta, P_\mathbf{X}*\Pi=P_\mathbf{Z}\}} E_{[P_\mathbf{X},\Pi]}\left[L_f(X^n,Z^n)\right] \right| = 0. \tag{23}$$

We defer the proof of Proposition 1 to the appendix.

As in Corollary 2, Proposition 1 gives a performance guarantee under the strict expectation criterion, i.e., when the maximization is over expectations rather than conditional expectations. It implies that under benign assumptions on the process, optimality with respect to the latter suffices for optimality with respect to the former.

# 7 Conclusion

In the discrete denoising problem, it is not always realistic to assume full knowledge of the channel characteristics. In this paper, we have presented a denoising scheme designed to operate in the presence of such channel uncertainty. We have proposed a worst case performance measure, argued its relevance for this setting, and established the universal asymptotic optimality of the suggested schemes under this criterion.

The schemes presented in this work can be practically implemented by identifying the problem of finding the minimizer in (1) with optimization problems that can be solved efficiently. The implementation aspects, along with experimental results on real and simulated data that seem to be indicative of the potential of these schemes to do well in practice, are presented in [8].

# Acknowledgment

Prof. Amir Dembo is gratefully acknowledged for helpful discussions.# Appendix

## A Technical Lemmas

In this section several technical lemmas are presented that are needed for the proofs of the main results. Before continuing, we define $\Lambda_{\max} = \max_{a,b} \Lambda(a,b)$.

The first lemma states that for any source and channel $\Pi$, $G_k\left(\hat{Q}^{2k+1}[Z^n], \Pi, f\right)$ is a very efficient estimate of $L_f(X^n,Z^n)$. In fact, it is uniformly efficient in all sources, channels, and sliding window functions $f$.

**Lemma 1** *For all $P_\mathbf{X} \in \mathcal{S}^\infty$, $\Pi \in \mathcal{Q}(\mathcal{A})$, $n > 2k$, $f \in \mathcal{F}_k$, and $\delta > 0$*

$$P_{[P_\mathbf{X},\Pi]}\left(\left|G_k\left(\hat{Q}^{2k+1}[Z^n],\Pi,f\right) - L_f(X^n,Z^n)\right| > \delta\right) \leq \exp\left[-nA\left(k,\delta,\Lambda_{\max},||\Pi^{-1}||\right)\right],$$

where $A(k, \delta, \Lambda_{\max}, ||\Pi^{-1}||)$ can be taken as any function satisfying

$$2(2k+1)|\mathcal{A}|^{2k+1} \exp\left(-\frac{2\delta^2(n-2k)}{(2k+1)|\mathcal{A}|^{4k+4}(\Lambda_{\max}||\Pi^{-1}||)^2}\right) \leq \exp\left[-nA\left(k, \delta, \Lambda_{\max}, ||\Pi^{-1}||\right)\right]. \quad (24)$$

*Remark:* We shall assume below that the $A$ chosen to satisfy (24) is non-decreasing in $\delta$ and non-increasing in $||\Pi^{-1}||$.

**Proof:**

We shall establish the Lemma by conditioning on the source sequence. Indeed, it will be enough to show that for all $P_\mathbf{X} \in \mathcal{S}^\infty$, $\Pi \in \mathcal{Q}(\mathcal{A})$, $f \in \mathcal{F}_k$, $\delta > 0$, and *all* $x^n \in \mathcal{A}^n$

$$P_{[P_\mathbf{X},\Pi]}\left(\left|G_k\left(\hat{Q}^{2k+1}[Z^n], \Pi, f\right) - L_f(x^n, Z^n)\right| > \delta \bigg| X^n = x^n\right) \leq$$
$$2(2k+1)|\mathcal{A}|^{2k+1} \exp\left(-\frac{2\delta^2(n-2k)}{(2k+1)|\mathcal{A}|^{4k+4}(\Lambda_{\max}||\Pi^{-1}||)^2}\right). \quad (25)$$

Note that when conditioning on $x^n$ in (25), $Z^n$ is a sequence of independent components, with $Z_i \sim \Pi(x_i, \cdot)$. Now

$$G_k\left(\hat{Q}^{2k+1}[Z^n], \Pi, f\right) = \sum_{z_{-k}^{-1}, z_1^k \in \mathcal{A}^k} \sum_{z \in \mathcal{A}} \mathbf{1}^T \left[\Pi^{-T} \hat{Q}^{2k+1}[Z^n]_{Z_0|z_{-k}^{-1}, z_1^k} \odot \boldsymbol{\pi}_z \odot \left[\Lambda \cdot f([z_{-k}^{-1}, z, z_1^k])\right]\right]$$

$$= \sum_{z_{-k}^{-1}, z_1^k \in \mathcal{A}^k} \sum_{z \in \mathcal{A}} \mathbf{1}^T \left[\Pi^{-T}\left(\frac{1}{n-2k}\sum_{i=k+1}^{n-k} \mathbf{1}_{\{Z_i = \cdot | z_{i-k}^{i-1}, z_{i+1}^{i+k}\}}\right) \odot \boldsymbol{\pi}_z \odot \left[\Lambda \cdot f([z_{-k}^{-1}, z, z_1^k])\right]\right]$$

$$= \frac{1}{n-2k}\sum_{i=k+1}^{n-k} \sum_{z_{-k}^{-1}, z_1^k \in \mathcal{A}^k} \sum_{z \in \mathcal{A}} \mathbf{1}^T \left[\Pi^{-T} \mathbf{1}_{\{Z_i = \cdot | z_{i-k}^{i-1}, z_{i+1}^{i+k}\}} \odot \boldsymbol{\pi}_z \odot \left[\Lambda \cdot f([z_{-k}^{-1}, z, z_1^k])\right]\right], \quad (26)$$

where:

- $\hat{Q}^{2k+1}[Z^n]_{Z_0|z_{-k}^{-1}, z_1^k}$ denotes the conditional distribution vector of $Z_0 | Z_{-k}^{-1} = z_{-k}^{-1}, Z_1^k = z_1^k$ induced by $\hat{Q}^{2k+1}[Z^n]$.

- $\mathbf{1}_{\{Z_i = \cdot | z_{i-k}^{i-1}, z_{i+1}^{i+k}\}}$ stands for the $|\mathcal{A}|$-dimensional column vector whose $a$-th component is zero unless $Z_{i-k}^{i+k} = (z_{i-k}^{i-1}, a, z_{i+1}^{i+k})$ in which case it is 1.

On the other hand,

$$L_f(x^n, Z^n) = \frac{1}{n-2k}\sum_{i=k+1}^{n-k} \sum_{a \in \mathcal{A}} \Lambda(x_i, a) f(Z_{i-k}^{i+k})[a]$$

$$= \frac{1}{n-2k}\sum_{i=k+1}^{n-k} [\Lambda \cdot f(Z_{i-k}^{i+k})]_{x_i}$$

$$= \frac{1}{n-2k}\sum_{i=k+1}^{n-k} \sum_{z_{-k}^{-1}, z_1^k \in \mathcal{A}^k} \sum_{z \in \mathcal{A}} \mathbf{1}^T \left[\mathbf{1}_{\{Z_{i-k}^{i+k} = (z_{-k}^{-1}, z, z_1^k), x_i = \cdot\}} \odot \left[\Lambda \cdot f([z_{-k}^{-1}, z, z_1^k])\right]\right], \quad (27)$$

where $\mathbf{1}_{\{Z_{i-k}^{i+k}=(z_{-k}^{-1},z,z_1^k),x_i=\cdot\}}$ denotes the $|\mathcal{A}|$-dimensional column vector whose $a$-th component is zero unless both $Z_{i-k}^{i+k} = (z_{-k}^{-1},z,z_1^k)$ and $x_i = a$ in which case it is 1. From (26), (27) and the triangle inequality it follows that

$$\left|G_k\left(\hat{Q}^{2k+1}[Z^n],\Pi,f\right) - L_f(x^n,Z^n)\right|$$

$$\leq |\mathcal{A}|\Lambda_{\max} \sum_{z_{-k}^{-1},z_1^k \in \mathcal{A}^k} \sum_{z \in \mathcal{A}} \left\| \frac{1}{n-2k} \sum_{i=k+1}^{n-k} \left(\mathbf{1}_{\{Z_{i-k}^{i+k}=(z_{-k}^{-1},z,z_1^k),x_i=\cdot\}} - \Pi^{-T}\mathbf{1}_{\{Z_i=\cdot|z_{i-k}^{i-1},z_{i+1}^{i+k}\}} \odot \boldsymbol{\pi}_z\right)\right\|_{\infty}$$

$$= |\mathcal{A}|\Lambda_{\max} \sum_{z_{-k}^{-1},z_1^k \in \mathcal{A}^k} \sum_{z \in \mathcal{A}} \max_{a \in \mathcal{A}} \left| \frac{1}{n-2k} \sum_{i=k+1}^{n-k} \left(\mathbf{1}_{\{Z_{i-k}^{i+k}=(z_{-k}^{-1},z,z_1^k),x_i=a\}} - \left[\Pi^{-T}\mathbf{1}_{\{Z_i=\cdot|z_{i-k}^{i-1},z_{i+1}^{i+k}\}} \odot \boldsymbol{\pi}_z\right](a)\right)\right|.$$
(28)

Now, for all $x^n \in \mathcal{A}^n$, contexts $z_{-k}^{-1}, z_1^k \in \mathcal{A}^k$, and $z \in \mathcal{A}$ we have,

$$P_{[P_\mathbf{X},\Pi]}\left(\left|\frac{1}{n-2k} \sum_{i=k+1}^{n-k} \left(\mathbf{1}_{\{Z_{i-k}^{i+k}=(z_{-k}^{-1},z,z_1^k),x_t=a\}} - \left[\Pi^{-T}\mathbf{1}_{\{Z_i=\cdot|z_{i-k}^{i-1},z_{i+1}^{i+k}\}} \odot \boldsymbol{\pi}_z\right](a)\right)\right| > \varepsilon \,\Big|\, X^n = x^n\right)$$

$$\leq 2(2k+1)\exp\left(-\frac{2\varepsilon^2(n-2k)}{(2k+1)(||\Pi^{-1}||)^2}\right).$$
(29)

We get (29) by decomposing the summation inside the probability on the left side of (29) into $2k+1$ sums of approximately $n/(2k+1)$ independent random variables bounded in magnitude by $||\Pi^{-1}||$, applying Hoeffding's inequality [10, Th. 1] to each of the sums, and combining via a union bound to obtain (29) (cf. similar derivations in [5] and [13]). Combining (28) and (29), with standard applications of the union bound, gives

$$P_{[P_\mathbf{X},\Pi]}\left(\left|G_k\left(\hat{Q}^{2k+1}[Z^n],\Pi,f\right) - L_f(x^n,Z^n)\right| > \delta \,\Big|\, X^n = x^n\right) \leq 2(2k+1)|\mathcal{A}|^{2k+1}\exp\left(-\frac{2\left(\frac{\delta}{\Lambda_{\max}|\mathcal{A}|^{2k+2}}\right)^2(n-2k)}{(2k+1)(||\Pi^{-1}||)^2}\right),$$

which, upon simplification of the expression in the exponent, is exactly (25).

$\square$

**Lemma 2** *For all $P_\mathbf{Z}$, and $P_\mathbf{X} \in \mathcal{S}^\infty$, $\Pi \in \mathcal{Q}(\mathcal{A})$ satisfying $P_\mathbf{X} * \Pi = P_\mathbf{Z}$,*

$$P_\mathbf{Z}\left(\left|G_k\left(\hat{Q}^{2k+1}[Z^n],\Pi,f\right) - E_{[P_\mathbf{X},\Pi]}\left[L_f(X^n,Z^n)|\mathbf{Z}\right]\right| > \delta\right) \leq \exp\left[-nB(k,\delta,\Lambda_{\max},||\Pi^{-1}||)\right]$$
(30)

*for all $n > 2k$, $f \in \mathcal{F}_k$, and $\delta > 0$, where $B(k,\delta,\Lambda_{\max},||\Pi^{-1}||)$ can be taken as any function satisfying*

$$\frac{|\mathcal{A}|^{2k+2}||\Pi^{-1}||\Lambda_{\max}}{\delta}\exp\left[-nA(k,\delta,\Lambda_{\max},||\Pi^{-1}||)\right] \leq \exp\left[-nB(k,\delta,\Lambda_{\max},||\Pi^{-1}||)\right].$$
(31)

*Remark:* Note that the random variables appearing in the probability on the left side of (30) are $\mathbf{Z}$-measurable, and hence it suffices to consider the probability measure $P_\mathbf{Z}$, which is the noisy marginal of $P_{[P_\mathbf{X},\Pi]}$. We shall assume below that the $B$ chosen to satisfy (31) is non-increasing in $||\Pi^{-1}||$. Finally, note that the combination of Lemma 1 and Lemma 2 implies that, for an *arbitrary* source $P_\mathbf{X}$ and channel $\Pi$, $E_{[P_\mathbf{X},\Pi]}[L_f(X^n,Z^n)|\mathbf{Z}] \approx L_f(X^n,Z^n)$ with high $P_{[P_\mathbf{X},\Pi]}$-probability.

**Proof:**

Fix $\varepsilon > 0$. By Lemma 1,

$$E_{P_\mathbf{Z}}\left[P_{[P_\mathbf{X},\Pi]}\left(\left|G_k\left(\hat{Q}^{2k+1}[Z^n],\Pi,f\right) - L_f(X^n,Z^n)\right| > \delta \Big| \mathbf{Z}\right)\right] \leq \exp\left[-nA(k,\delta,\Lambda_{\max},||\Pi^{-1}||)\right],$$

implying, by Chebyshev's inequality,

$$P_\mathbf{Z}\left(P_{[P_\mathbf{X},\Pi]}\left(\left|G_k\left(\hat{Q}^{2k+1}[Z^n],\Pi,f\right) - L_f(X^n,Z^n)\right| > \delta \Big| \mathbf{Z}\right) > \varepsilon\right) \leq \frac{1}{\varepsilon}\exp\left[-nA(k,\delta,\Lambda_{\max},||\Pi^{-1}||)\right]. \quad (32)$$

Now, the fact that $\left|G_k\left(\hat{Q}^{2k+1}[Z^n],\Pi,f\right) - L_f(X^n,Z^n)\right| \leq |\mathcal{A}|^{2k+2}||\Pi^{-1}||\Lambda_{\max}$ implies that

$$\left|G_k\left(\hat{Q}^{2k+1}[Z^n],\Pi,f\right) - E_{[P_\mathbf{X},\Pi]}\left[L_f(X^n,Z^n)|\mathbf{Z}\right]\right| \leq \delta + \varepsilon|\mathcal{A}|^{2k+2}||\Pi^{-1}||\Lambda_{\max}$$

on the event

$$\left\{P_{[P_\mathbf{X},\Pi]}\left(\left|G_k\left(\hat{Q}^{2k+1}[Z^n],\Pi,f\right) - L_f(X^n,Z^n)\right| > \delta \Big| \mathbf{Z}\right) \leq \varepsilon\right\},$$

in turn implying

$$P_\mathbf{Z}\left(\left|G_k\left(\hat{Q}^{2k+1}[Z^n],\Pi,f\right) - E_{[P_\mathbf{X},\Pi]}\left[L_f(X^n,Z^n)|\mathbf{Z}\right]\right| > \delta + \varepsilon|\mathcal{A}|^{2k+2}||\Pi^{-1}||\Lambda_{\max}\right) \leq \frac{1}{\varepsilon}\exp\left[-nA(k,\delta,\Lambda_{\max},||\Pi^{-1}||)\right]$$

when combined with (32). Choosing $\varepsilon$ such that $\delta = \varepsilon|\mathcal{A}|^{2k+2}||\Pi^{-1}||\Lambda_{\max}$, this implies

$$P_\mathbf{Z}\left(\left|G_k\left(\hat{Q}^{2k+1}[Z^n],\Pi,f\right) - E_{[P_\mathbf{X},\Pi]}\left[L_f(X^n,Z^n)|\mathbf{Z}\right]\right| > 2\delta\right) \leq \frac{|\mathcal{A}|^{2k+2}||\Pi^{-1}||\Lambda_{\max}}{\delta}\exp\left[-nA(k,\delta,\Lambda_{\max},||\Pi^{-1}||)\right],$$

from which an explicit form for the exponent function $B$ in the right side of (30) can be obtained.

$\square$

The next lemma states that, with high probability, $G_k\left(\hat{Q}^{2k+1}[Z^n],\Pi,f\right)$ estimates $E_{[P_\mathbf{X},\Pi]}\left[L_f(X^n,Z^n)|\mathbf{Z}\right]$ uniformly well, simultaneously for all $f \in \mathcal{F}_k$ and any finite number of pairs $(P_\mathbf{X},\Pi)$ that give rise to $P_\mathbf{Z}$.

**Lemma 3** *For all $P_\mathbf{Z} \in \mathcal{S}^\infty$, finite $\mathcal{K} \subseteq \mathcal{C}_\infty(P_\mathbf{Z})$,*

$$P_\mathbf{Z}\left(\max_{f \in \mathcal{F}_k} \sup_{\{(P_\mathbf{X},\Pi):\Pi \in \mathcal{K}, P_\mathbf{X}*\Pi=P_\mathbf{Z}\}} \left|G_k\left(\hat{Q}^{2k+1}[Z^n],\Pi,f\right) - E_{[P_\mathbf{X},\Pi]}\left[L_f(X^n,Z^n)|\mathbf{Z}\right]\right| > \eta + \delta\right)$$

$$\leq \left[\frac{\Lambda_{\max}\left(1 + |\mathcal{A}|^{2k+2}\max_{\Pi \in \mathcal{K}}||\Pi^{-1}||\right)}{\eta}\right]^{|\mathcal{A}|^{2k+2}} \cdot |\mathcal{K}| \cdot \exp\left[-nB\left(k,\delta,\Lambda_{\max},\max_{\Pi \in \mathcal{K}}||\Pi^{-1}||\right)\right] \quad (33)$$

*for all $n > 2k$ and $\delta, \eta > 0$.*

**Proof:**

Lemma 2, the union bound, and the fact that $B(k,\delta,\Lambda_{\max},\cdot)$ is non-increasing imply that for any $f \in \mathcal{F}_k$

$$P_\mathbf{Z}\left(\sup_{\{(P_\mathbf{X},\Pi):\Pi \in \mathcal{K}, P_\mathbf{X}*\Pi=P_\mathbf{Z}\}} \left|G_k\left(\hat{Q}^{2k+1}[Z^n],\Pi,f\right) - E_{[P_\mathbf{X},\Pi]}\left[L_f(X^n,Z^n)|\mathbf{Z}\right]\right| > \delta\right)$$

$$\leq |\mathcal{K}|\exp\left[-nB\left(k,\delta,\Lambda_{\max},\max_{\Pi \in \mathcal{K}}||\Pi^{-1}||\right)\right]. \quad (34)$$

For $\varepsilon > 0$ let $\mathcal{S}(\mathcal{A}, \varepsilon)$ denote the subset of $\mathcal{S}(\mathcal{A})$ consisting of distributions that assign probabilities that are integer multiples of $\varepsilon$ to each $a \in \mathcal{A}$. Letting $\mathcal{F}_k^\varepsilon \triangleq \{f : \mathcal{A}^{2k+1} \to \mathcal{S}_\varepsilon(\mathcal{A})\}$, it is then straightforward from the definition of $G_k$ and of $L_f$ that

$$\max_{f \in \mathcal{F}_k} \sup_{\{(P_\mathbf{X}, \Pi): \Pi \in \mathcal{K}, P_\mathbf{X} * \Pi = P_\mathbf{Z}\}} \left| G_k \left( \hat{Q}^{2k+1}[Z^n], \Pi, f \right) - E_{[P_\mathbf{X}, \Pi]} \left[ L_f(X^n, Z^n) | \mathbf{Z} \right] \right|$$
$$\leq \max_{f \in \mathcal{F}_k^\varepsilon} \sup_{\{(P_\mathbf{X}, \Pi): \Pi \in \mathcal{K}, P_\mathbf{X} * \Pi = P_\mathbf{Z}\}} \left| G_k \left( \hat{Q}^{2k+1}[Z^n], \Pi, f \right) - E_{[P_\mathbf{X}, \Pi]} \left[ L_f(X^n, Z^n) | \mathbf{Z} \right] \right|$$
$$+ \varepsilon \Lambda_{\max} \left( 1 + |\mathcal{A}|^{2k+2} \max_{\Pi \in \mathcal{K}} ||\Pi^{-1}|| \right). \tag{35}$$

Combining (34), (35), and the fact that $|\mathcal{F}_k^\varepsilon| = |\mathcal{S}(\mathcal{A}, \varepsilon)|^{|\mathcal{A}|^{2k+1}} \leq \left( \frac{1}{\varepsilon |\mathcal{A}|} \right)^{|\mathcal{A}|^{2k+1}} = \varepsilon^{-|\mathcal{A}|^{2k+2}}$ yields, for $\eta = \varepsilon \Lambda_{\max} \left( 1 + |\mathcal{A}|^{2k+2} \max_{\Pi \in \mathcal{K}} ||\Pi^{-1}|| \right)$,

$$P_\mathbf{Z} \left( \max_{f \in \mathcal{F}_k} \sup_{\{(P_\mathbf{X}, \Pi): \Pi \in \mathcal{K}, P_\mathbf{X} * \Pi = P_\mathbf{Z}\}} \left| G_k \left( \hat{Q}^{2k+1}[Z^n], \Pi, f \right) - E_{[P_\mathbf{X}, \Pi]} \left[ L_f(X^n, Z^n) | \mathbf{Z} \right] \right| > \eta + \delta \right)$$
$$\leq |\mathcal{F}_k^\varepsilon| |\mathcal{K}| \exp \left[ -nB \left( k, \delta, \Lambda_{\max}, \max_{\Pi \in \mathcal{K}} ||\Pi^{-1}|| \right) \right]$$
$$\leq \varepsilon^{-|\mathcal{A}|^{2k+2}} |\mathcal{K}| \exp \left[ -nB \left( k, \delta, \Lambda_{\max}, \max_{\Pi \in \mathcal{K}} ||\Pi^{-1}|| \right) \right],$$

which is exactly (33) since $\frac{1}{\varepsilon} = \frac{\Lambda_{\max} \left( 1 + |\mathcal{A}|^{2k+2} \max_{\Pi \in \mathcal{K}} ||\Pi^{-1}|| \right)}{\eta}$.

□

**Lemma 4** *For all* $P_\mathbf{Z} \in \mathcal{S}^\infty$, *finite* $\mathcal{K} \subseteq \mathcal{C}_\infty(P_\mathbf{Z})$,

$$P_\mathbf{Z} \left( \max_{f \in \mathcal{F}_k} \sup_{\{(P_\mathbf{X}, \Pi): \Pi \in \mathcal{K}, P_\mathbf{X} * \Pi = P_\mathbf{Z}\}} \left| G_k \left( \hat{Q}^{2k+1}[Z^n], \Pi, f \right) - E_{[P_\mathbf{X}, \Pi]} \left[ L_f(X^n, Z^n) | \mathbf{Z} \right] \right| > \delta \right)$$
$$\leq |\mathcal{K}| \cdot \exp \left[ -n\Gamma \left( k, \delta, \Lambda_{\max}, \max_{\Pi \in \mathcal{K}} ||\Pi^{-1}|| \right) \right]$$

*for all* $n > 2k$ *and* $\delta > 0$, *where* $\Gamma$ *can be any function satisfying*

$$\left[ \frac{2\Lambda_{\max} \left( 1 + |\mathcal{A}|^{2k+2} \max_{\Pi \in \mathcal{K}} ||\Pi^{-1}|| \right)}{\delta} \right]^{|\mathcal{A}|^{2k+2}} \exp \left[ -nB \left( k, \delta/2, \Lambda_{\max}, \max_{\Pi \in \mathcal{K}} ||\Pi^{-1}|| \right) \right] \leq \exp \left[ -n\Gamma \left( k, \delta, \Lambda_{\max}, \max_{\Pi \in \mathcal{K}} ||\Pi^{-1}|| \right) \right]$$

**Proof:**

The assertion follows from Lemma 3 upon assigning $\delta' = \delta/2$, $\eta = \delta/2$, and noting the decreasing monotonicity of $B(k, \delta, \Lambda_{\max}, \cdot)$, with $\Gamma$ chosen to be any function satisfying

$$\left[ \frac{2\Lambda_{\max} \left( 1 + |\mathcal{A}|^{2k+2} \max_{\Pi \in \mathcal{K}} ||\Pi^{-1}|| \right)}{\delta} \right]^{|\mathcal{A}|^{2k+2}} \exp \left[ -nB \left( k, \delta/2, \Lambda_{\max}, \max_{\Pi \in \mathcal{K}} ||\Pi^{-1}|| \right) \right] \leq \exp \left[ -n\Gamma \left( k, \delta, \Lambda_{\max}, \max_{\Pi \in \mathcal{K}} ||\Pi^{-1}|| \right) \right] \tag{36}$$

□

Note that in Lemmas 2, 3, and 4, $P_\mathbf{Z}$ is a completely arbitrary distribution, which need not even be stationary.

We now define $\hat{\Delta}_l(\hat{Q}^{2l+1}[Z^n], \Delta) = \Delta \cap \mathcal{C}_l(\hat{Q}^{2l+1}[Z^n])$.[11] The denoiser in Section 3.B is defined as a function of $\hat{\Delta}_l(\hat{Q}^{2l+1}[Z^n], \Delta)$, as opposed to $\Delta \cap \mathcal{C}_\infty(P_\mathbf{Z})$ which would be ideal. Clearly this is not possible since $P_\mathbf{Z}$ is not known. However we expect that $\hat{\Delta}_l$ will be close to $\Delta \cap \mathcal{C}_\infty(P_\mathbf{Z})$. This is indeed the case, as quantified in Lemmas 5 and 6 below.

Before we state our final three lemmas, we need to set up some notation. Denote by $\tilde{\mathcal{S}} \subset \mathcal{S}^\infty$, the set of stationary distribution in $\mathcal{S}^\infty$. Further, for $\alpha = \alpha(n, l, \varepsilon)$, let $\tilde{\mathcal{S}}(\mathcal{A}, \alpha)$ denote the set of all $P_\mathbf{Z} \in \tilde{\mathcal{S}}$ for which

$$P_\mathbf{Z}\left(\left\|P_{Z^l_{-l}} - \hat{Q}^{2l+1}[Z^n]\right\| > \varepsilon\right) \leq \alpha(n, l, \varepsilon)$$

holds for all $n, l, \varepsilon$. Note that by the Borel–Cantelli lemma, for $\alpha$ satisfying $\sum_n \alpha(n, l, \varepsilon) < \infty$ for all $l$ and $\varepsilon > 0$, $\tilde{\mathcal{S}}(\mathcal{A}, \alpha)$ is a subset of the stationary and ergodic sources. For any $P_\mathbf{Z} \in \tilde{\mathcal{S}}$ and uncertainty set $\Delta$, let

$$a_l = a_l(P_\mathbf{Z}, \Delta) = \rho\left(\Delta \cap \mathcal{C}_\infty(P_\mathbf{Z}), \Delta \cap \mathcal{C}_l(P_\mathbf{Z})\right). \tag{37}$$

For a given $\alpha$ define now

$$U_n(k, l, \eta, \delta) = \alpha(n, l, b_l^{-1}(\phi_k^{-1}(\delta - \phi_k(\eta)) - a_l)), \tag{38}$$

where $\phi_k$ is defined in (9) and $b_l$ is the function associated with Assumption 2. Let further

$$V_n(k, l, \delta) = U_n(k, l, \exp(-\sqrt{n}/|\mathcal{A}|^2), \delta). \tag{39}$$

**Lemma 5** *For all $P_\mathbf{Z} \in \tilde{\mathcal{S}}$ and $\Delta \subseteq \mathcal{Q}(\mathcal{A})$ with $\max_{\Pi \in \Delta} ||\Pi^{-1}|| < \infty$,*

$$P_\mathbf{Z}\left(\rho\left(\Delta \cap \mathcal{C}_\infty(P_\mathbf{Z}), \hat{\Delta}_l\left(\hat{Q}^{2l+1}[Z^n]\right)\right) > \delta\right) \leq P_\mathbf{Z}\left(\left\|P_{Z^l_{-l}} - \hat{Q}^{2l+1}[Z^n]\right\| > b_l^{-1}(\delta - a_l)\right),$$

*where $b_l$ and $a_l$ were defined in (20) and (37), respectively.*

**Proof:**

We have

$$P_\mathbf{Z}\left(\rho\left(\Delta \cap \mathcal{C}_\infty(P_\mathbf{Z}), \hat{\Delta}_l\left(\hat{Q}^{2l+1}[Z^n], \Delta\right)\right) > \delta\right) \leq P_\mathbf{Z}\left(\rho\left(\Delta \cap \mathcal{C}_l(P_\mathbf{Z}), \hat{\Delta}_l\left(\hat{Q}^{2l+1}[Z^n], \Delta\right)\right) > \delta - a_l\right)$$
$$\leq P_\mathbf{Z}\left(\left\|P_{Z^l_{-l}} - \hat{Q}^{2l+1}[Z^n]\right\| > b_l^{-1}(\delta - a_l)\right)$$

where the first inequality follows by the definition of $a_l$, as defined in (37), and the triangle inequality, and the second inequality follows from the definition of $b_l$, as defined in (20), and the definition of $\hat{\Delta}_l$.

□

**Lemma 6** *For any $P_\mathbf{Z} \in \tilde{\mathcal{S}}$ and $\Delta$ satisfying Assumption 1, the sequence $a_l(P_\mathbf{Z}, \Delta)$ defined in (37) satisfies $a_l(P_\mathbf{Z}, \Delta) \to 0$ as $l \to \infty$. Furthermore, the convergence is uniform in $P_\mathbf{Z}$.*

---
[11] We may suppress $\hat{\Delta}_l$ dependence on $\Delta$ and $\hat{Q}^{2l+1}[Z^n]$.

**Proof:**

The first thing to establish is that the relation

$$\Delta \cap \mathcal{C}_\infty(P_\mathbf{Z}) = \bigcap_{k \geq 1} (\Delta \cap \mathcal{C}_k(P_\mathbf{Z})) = \bigcap_{k \geq 1} \left\{ \Pi \in \Delta : \exists P_{X_{-k}^k} \in \mathcal{S}^{2k+1} \text{ s.t. } P_{X_{-k}^k} * \Pi = P_{Z_{-k}^k} \right\} \quad (40)$$

holds for all stationary $P_\mathbf{Z}$. The direction $\subseteq$ is true since obviously

$$\Delta \cap \mathcal{C}_\infty(P_\mathbf{Z}) \subseteq \left\{ \Pi \in \Delta : \exists P_{X_{-k}^k} \in \mathcal{S}^{2k+1} \text{ s.t. } P_{X_{-k}^k} * \Pi = P_{Z_{-k}^k} \right\}$$

for every $k$. For the reverse direction note that if $P'_{X_{-k}^k} = P_{Z_{-k}^k} *^{-1} \Pi \in \mathcal{S}^{2k+1}$ and $P'_{X_{-(k+1)}^{k+1}} = P_{Z_{-(k+1)}^{k+1}} *^{-1} \Pi \in \mathcal{S}^{2k+3}$ then $P'_{X_{-k}^k}$ is consistent with $P'_{X_{-(k+1)}^{k+1}}$, i.e. its $2k+1$-th order marginal. Thus, if $\Pi$ is in the intersection of the sets on the right side of (40) then $\{P'_{X_{-k}^k}\}_{k \geq 1}$ is a consistent family of distributions so, by Kolmogorov's extension theorem, there exists an unique stationary source $P'_\mathbf{X}$ with the said distributions as its finite-dimensional marginals. Furthermore, $P'_\mathbf{X} * \Pi = P_\mathbf{Z}$ since $P'_{X_{-k}^k} * \Pi = P_{Z_{-k}^k}$ for each $k$. Thus we have

$$\Delta \cap \mathcal{C}_\infty(P_\mathbf{Z}) \supseteq \left\{ \Pi \in \Delta : \exists P_{X_{-k}^k} \in \mathcal{S}^{2k+1} \text{ s.t. } P_{X_{-k}^k} * \Pi = P_{Z_{-k}^k} \right\},$$

establishing (40). Now, the fact that $\{\Delta \cap \mathcal{C}_k(P_\mathbf{Z})\}$ is a decreasing sequence and that $\Delta \cap \mathcal{C}_\infty(P_\mathbf{Z}) \subseteq \Delta \cap \mathcal{C}_k(P_\mathbf{Z})$ for all $k$ implies existence of the limit $\lim_{k \to \infty} \rho(\Delta \cap \mathcal{C}_\infty(P_\mathbf{Z}), \Delta \cap \mathcal{C}_k(P_\mathbf{Z}))$. Assume

$$\lim_{k \to \infty} \rho(\Delta \cap \mathcal{C}_\infty(P_\mathbf{Z}), \Delta \cap \mathcal{C}_k(P_\mathbf{Z})) > 0. \quad (41)$$

Let

$$\gamma = \lim_{k \to \infty} \rho(\Delta \cap \mathcal{C}_\infty(P_\mathbf{Z}), \Delta \cap \mathcal{C}_k(P_\mathbf{Z})) > 0$$

and define

$$\mathcal{C}_k^\gamma(P_\mathbf{Z}) = \left\{ \pi \in (\Delta \cap \mathcal{C}_k(P_\mathbf{Z}))^- \ s.t. \inf_{\pi' \in \Delta \cap \mathcal{C}_\infty(P_\mathbf{Z})} ||\pi - \pi'|| \geq \frac{\gamma}{2} \right\}$$

here we use the notation $A^-$ for the closure of set $A$. Be definition of $\gamma$, $\mathcal{C}_k^\gamma(P_\mathbf{Z}) \neq \emptyset$ for all $k$ and since $\mathcal{C}_k(P_\mathbf{Z}) \subseteq \mathcal{C}_{k-1}(P_\mathbf{Z})$ then $\mathcal{C}_k^\gamma(P_\mathbf{Z}) \subseteq \mathcal{C}_{k-1}^\gamma(P_\mathbf{Z})$ for all $k$. We also observe that $\mathcal{C}_k^\gamma(P_\mathbf{Z})$ is closed for all $k$. This last step follows from the fact that our norm $|| \cdot ||$ agrees with the given topology. By Assumption 1 and (40) we have $\cap_{k=1}^\infty \mathcal{C}_k^\gamma(P_\mathbf{Z}) \subseteq (\Delta \cap \mathcal{C}_\infty(P_\mathbf{Z}))^-$. Since $\{\mathcal{C}_k^\gamma(P_\mathbf{Z})\}$ is a nested sequence of closed and bounded sets, the bounding comes from the fact that the set of all channels is itself a bounded set, there exists $\pi \in \cap_{k=1}^\infty \mathcal{C}_k^\gamma(P_\mathbf{Z}) \subseteq (\Delta \cap \mathcal{C}_\infty(P_\mathbf{Z}))^-$. This would mean that

$$\inf_{\Pi' \in \Delta \cap \mathcal{C}_\infty(P_\mathbf{Z})} ||\pi - \Pi'|| \geq \frac{\gamma}{2}$$

which is false since $\pi \in (\Delta \cap \mathcal{C}_\infty(P_\mathbf{Z}))^-$ and $|| \cdot ||$ is a continuous function. Hence (41) is wrong and

$$\lim_{k \to \infty} \rho(\Delta \cap \mathcal{C}_\infty(P_\mathbf{Z}), \Delta \cap \mathcal{C}_k(P_\mathbf{Z})) = 0.$$

Therefore, for each $P_\mathbf{Z}$, $\lim_{l \to \infty} a_l(P_\mathbf{Z}, \Delta) = 0$. Since the set of distributions $\tilde{\mathcal{S}}$ is compact and from Assumption 1 we know

$$\rho(\Delta \cap \mathcal{C}_\infty(P_\mathbf{Z}), \Delta \cap \mathcal{C}_k(P_\mathbf{Z}))$$

is continuous in $P_\mathbf{Z}$, Dini's Theorem implies the convergence is uniform in $P_\mathbf{Z}$.

□

We can now state a generalized version of Theorem 2.

**Lemma 7** *For any $P_\mathbf{Z} \in \tilde{\mathcal{S}}_\alpha$, let*

$$\hat{X}^n_{univ} = \hat{X}^{n,k_n,l_n}_\Delta, \tag{42}$$

*where on the right side is the n-block denoiser defined in (3) and $\{k_n\}$, $\{l_n\}$ are unbounded increasing sequences satisfying $k_n \leq \frac{\ln n}{16 \ln |\mathcal{A}|}$ and $\sum_n V_n(k_n, l_n, \delta) < \infty$ for every $\delta > 0$. If $\mathcal{C}_\infty(P_\mathbf{Z}) \cap \Delta \neq \emptyset$ and sequence $\{\Delta_n\}$ with $\Delta_n \subseteq \Delta$ and $|\Delta_n| = O\left(e^{\sqrt{n}}\right)$, then*

$$\limsup_{n \to \infty} \left[ \mathcal{L}_{\hat{X}^n_{univ}}(P_\mathbf{Z}, \Delta_n, \mathbf{Z}) - \mu^{(n)}_{k_n}(P_\mathbf{Z}, \Delta, \mathbf{Z}) \right] \leq 0 \qquad P_\mathbf{Z} - a.s. \tag{43}$$

*Remarks:*

- The extreme detail of Lemma 7 makes it hard to extract any intuition from it. The main purpose of the lemma is to develop the subsequent Theorem 2 and Proposition 1.

- Note that the stipulation in the statement of the theorem that $\mathcal{C}_\infty(P_\mathbf{Z}) \cap \Delta \neq \emptyset$ is not restrictive since the real channel is known to lie in $\Delta$.

- To avoid introducing additional notation, henceforth $\hat{X}^n_{univ}$ denotes the denoiser defined in (42), rather than that of Theorem 1.

- It should be emphasized that the sequence $\{\Delta_n\}$ is not related to the construction of the denoiser. Rather, $\Delta_n$ is simply the subset of $\Delta$ on which performance is evaluated for the $n$-block denoiser (cf. (43)). Note that since the size of $\Delta_n$ is allowed to grow quite rapidly, one can choose a sequence $\{\Delta_n\}$ for which $\rho(\Delta_n, \Delta) \to 0$ quickly.

**Proof:**

We start by outlining the proof idea. Two ingredients that were absent in the setting of Theorem 1 and that now need to be accommodated are the fact that $\Delta$ is not necessarily finite, and that $\Delta$ need not be a subset of $\mathcal{C}_\infty(P_\mathbf{Z})$. The first ingredient is accommodated by evaluating performance, for each $n$, on a finite subset of $\Delta$, $\Delta_n$. For the second ingredient noted, a good thing to do would have been to employ the denoiser $\hat{X}^{n,k}_{\Delta'}$ taking $\Delta' = \Delta \cap \mathcal{C}_\infty(P_\mathbf{Z})$. Instead, the denoiser we construct in the present theorem is $\hat{X}^{n,k,l}_\Delta$. Lemmas 5 and 6 ensure that for large enough $l$, $\hat{\Delta}_l$ is "close" to $\Delta \cap \mathcal{C}_\infty(P_\mathbf{Z})$ which, in turn, implies that the performance of the scheme that uses $\hat{\Delta}_l$ is essentially as good as one which would be based on $\Delta \cap \mathcal{C}_\infty(P_\mathbf{Z})$. The bounds in the lemmas, when combined with the additional stipulation of Lemma 7, that $P_\mathbf{Z} \in \tilde{\mathcal{S}}(\mathcal{A}, \alpha)$ provide growth rates for $k$ and $l$ which guarantee that under the $\rho$ metric, $\hat{\Delta}_l\left(\hat{Q}^{2l+1}[Z^n]\right) \to \Delta \cap \mathcal{C}_\infty(P_\mathbf{Z})$ rapidly enough to ensure that the performance of $\hat{X}^{n,k_n,l_n}_\Delta$ converges to the performance of $\hat{X}^{n,k_n}_{\Delta \cap \mathcal{C}_\infty(P_\mathbf{Z})}$. It should be noted that the only point where the stationarity and mixing conditions, on the noise-corrupted source are used is for the estimation of $\Delta \cap \mathcal{C}_\infty(P_\mathbf{Z})$. For a completely arbitrary $P_\mathbf{Z}$, not necessarily stationary, if $\Delta \cap \mathcal{C}_\infty(P_\mathbf{Z})$ were given then the scheme of Theorem 1, where $\Delta \cap \mathcal{C}_\infty(P_\mathbf{Z})$ is used for $\Delta$, could be used, and the performance guarantees of Theorem 1 would apply. In the remainder of this subsection we give the rigorous proof of Lemma 7.

Lemma 6 and the fact that $P_\mathbf{Z} \in \tilde{\mathcal{S}}(\mathcal{A}, \alpha)$ imply (recall (20) and (37) for definitions of $b_l$ and $a_l$)

$$P_\mathbf{Z}\left(\rho\left(\Delta \cap \mathcal{C}_\infty(P_\mathbf{Z}), \hat{\Delta}_l\left(\hat{Q}^{2l+1}[Z^n], \Delta\right)\right) > \delta\right) \leq \alpha(n, l, b_l^{-1}(\delta - a_l)). \quad (44)$$

Combined with (9) this implies

$$\begin{aligned}
& P_\mathbf{Z}\left(\max_{f \in \mathcal{F}_k}\left|J_k\left(\hat{Q}^{2k+1}[Z^n], \Delta \cap \mathcal{C}_\infty(P_\mathbf{Z}), f\right) - J_k\left(\hat{Q}^{2k+1}[Z^n], \hat{\Delta}_l\left(\hat{Q}^{2l+1}[Z^n], \Delta\right), f\right)\right| > \delta\right) \\
& \leq P_\mathbf{Z}\left(\rho\left(\Delta \cap \mathcal{C}_\infty(P_\mathbf{Z}), \hat{\Delta}_l\left(\hat{Q}^{2l+1}[Z^n], \Delta\right)\right) > \phi_k^{-1}(\delta)\right) \\
& \leq \alpha(n, l, b_l^{-1}(\phi_k^{-1}(\delta) - a_l)). \quad (45)
\end{aligned}$$

Let now $\Delta_{[\eta]}$ denote an $\eta$-cover of $\Delta$. Note that for all sample paths, by (9) and the fact that $\Delta_n \subseteq \Delta$ implies $\rho(\Delta_n \cup \Delta_{[\eta]}, \Delta) \leq \eta$,

$$\max_{f \in \mathcal{F}_k}\left|J_k\left(\hat{Q}^{2k+1}[Z^n], (\Delta_n \cup \Delta_{[\eta]}) \cap \mathcal{C}_\infty(P_\mathbf{Z}), f\right) - J_k\left(\hat{Q}^{2k+1}[Z^n], \Delta \cap \mathcal{C}_\infty(P_\mathbf{Z}), f\right)\right| \leq \phi_k(\eta). \quad (46)$$

The combination of (46) with (45) now implies

$$P_\mathbf{Z}\left(\max_{f \in \mathcal{F}_k}\left|J_k\left(\hat{Q}^{2k+1}[Z^n], (\Delta_n \cup \Delta_{[\eta]}) \cap \mathcal{C}_\infty(P_\mathbf{Z}), f\right) - J_k\left(\hat{Q}^{2k+1}[Z^n], \hat{\Delta}_l\left(\hat{Q}^{2l+1}[Z^n], \Delta\right), f\right)\right| > \delta\right) \leq U_n(k, l, \eta, \delta), \quad (47)$$

where $U_n$ was defined in (38). Now, from the definition of $\hat{X}_\Delta^{n,k,l}$, it follows that

$$\begin{aligned}
& \sup_{\{(P_\mathbf{X}, \Pi): \Pi \in \Delta_n \cup \Delta_{[\eta]}, P_\mathbf{X} * \Pi = P_\mathbf{Z}\}} E_{[P_\mathbf{X}, \Pi]}\left[L_{\hat{X}_\Delta^{n,k,l}}(X^n, Z^n) | \mathbf{Z}\right] \\
& = \sup_{\{(P_\mathbf{X}, \Pi): \Pi \in \Delta_n \cup \Delta_{[\eta]}, P_\mathbf{X} * \Pi = P_\mathbf{Z}\}} E_{[P_\mathbf{X}, \Pi]}\left[L_{f_{\mathrm{MM}_k}[\hat{Q}^{2k+1}[Z^n], \hat{\Delta}_l(\hat{Q}^{2l+1}[Z^n], \Delta)]}(X^n, Z^n) | \mathbf{Z}\right]. \quad (48)
\end{aligned}$$

On the other hand, for every $f \in \mathcal{F}_k$,

$$\begin{aligned}
\sup_{\{(P_\mathbf{X}, \Pi): \Pi \in \Delta_n \cup \Delta_{[\eta]}, P_\mathbf{X} * \Pi = P_\mathbf{Z}\}} G_k\left(\hat{Q}^{2k+1}[Z^n], \Pi, f\right) & = \max_{\Pi \in (\Delta_n \cup \Delta_{[\eta]}) \cap \mathcal{C}_\infty(P_\mathbf{Z})} G_k\left(\hat{Q}^{2k+1}[Z^n], \Pi, f\right) \\
& = J_k\left(\hat{Q}^{2k+1}[Z^n], (\Delta_n \cup \Delta_{[\eta]}) \cap \mathcal{C}_\infty(P_\mathbf{Z}), f\right) \quad (49)
\end{aligned}$$

implying, when combined with Lemma 4, that

$$\begin{aligned}
& P_\mathbf{Z}\left(\max_{f \in \mathcal{F}_k}\left|J_k\left(\hat{Q}^{2k+1}[Z^n], (\Delta_n \cup \Delta_{[\eta]}) \cap \mathcal{C}_\infty(P_\mathbf{Z}), f\right) - \sup_{\{(P_\mathbf{X}, \Pi): \Pi \in \Delta_n \cup \Delta_{[\eta]}, P_\mathbf{X} * \Pi = P_\mathbf{Z}\}} E_{[P_\mathbf{X}, \Pi]}\left[L_f(X^n, Z^n) | \mathbf{Z}\right]\right| > \delta\right) \\
& \leq (|\Delta_n| + |\Delta_{[\eta]}|) \cdot \exp\left[-n\Gamma\left(k, \delta, \Lambda_{\max}, \max_{\Pi \in \Delta}||\Pi^{-1}||\right)\right]. \quad (50)
\end{aligned}$$

When (50) is combined with (47) as well as a union bound and a triangle inequality, we get

$$\begin{aligned}
& P_\mathbf{Z}\left(\max_{f \in \mathcal{F}_k}\left|J_k\left(\hat{Q}^{2k+1}[Z^n], \hat{\Delta}_l\left(\hat{Q}^{2l+1}[Z^n], \Delta\right), f\right) - \sup_{\{(P_\mathbf{X}, \Pi): \Pi \in \Delta_n \cup \Delta_{[\eta]}, P_\mathbf{X} * \Pi = P_\mathbf{Z}\}} E_{[P_\mathbf{X}, \Pi]}\left[L_f(X^n, Z^n) | \mathbf{Z}\right]\right| > \delta\right) \\
& \leq U_n(k, l, \eta, \delta/2) + (|\Delta_n| + |\Delta_{[\eta]}|) \cdot \exp\left[-n\Gamma\left(k, \delta/2, \Lambda_{\max}, \max_{\Pi \in \Delta}||\Pi^{-1}||\right)\right]. \quad (51)
\end{aligned}$$

Since by the definition of $f_{\text{MM}_k}$

$$J_k\left(\hat{Q}^{2k+1}[Z^n], \hat{\Delta}_l\left(\hat{Q}^{2l+1}[Z^n], \Delta\right), f_{\text{MM}_k}\left[\hat{Q}^{2k+1}[Z^n], \hat{\Delta}_l\left(\hat{Q}^{2l+1}[Z^n], \Delta\right)\right]\right)$$
$$= \min_{f \in \mathcal{F}_k} J_k\left(\hat{Q}^{2k+1}[Z^n], \hat{\Delta}_l\left(\hat{Q}^{2l+1}[Z^n], \Delta\right), f\right), \tag{52}$$

it follows that

$$P_{\mathbf{Z}}\left(\left|\min_{f \in \mathcal{F}_k} J_k\left(\hat{Q}^{2k+1}[Z^n], \hat{\Delta}_l\left(\hat{Q}^{2l+1}[Z^n], \Delta\right), f\right) - \sup_{\{(P_{\mathbf{X}}, \Pi): \Pi \in \Delta_n \cup \Delta_{[\eta]}, P_{\mathbf{X}} * \Pi = P_{\mathbf{Z}}\}} E_{[P_{\mathbf{X}}, \Pi]}\left[L_{\hat{X}_{\Delta}^{n,k,l}}(X^n, Z^n) | \mathbf{Z}\right]\right| > \delta\right)$$
$$= P_{\mathbf{Z}}\left(\left|J_k\left(\hat{Q}^{2k+1}[Z^n], \hat{\Delta}_l\left(\hat{Q}^{2l+1}[Z^n], \Delta\right), f_{\text{MM}_k}\right) - \sup_{\{(P_{\mathbf{X}}, \Pi): \Pi \in \Delta_n \cup \Delta_{[\eta]}, P_{\mathbf{X}} * \Pi = P_{\mathbf{Z}}\}} E_{[P_{\mathbf{X}}, \Pi]}\left[L_{f_{\text{MM}_k}}(X^n, Z^n) | \mathbf{Z}\right]\right| > \delta\right)$$
$$\leq U_n(k, l, \eta, \delta/2) + \left(|\Delta_n| + |\Delta_{[\eta]}|\right) \cdot \exp\left[-n\Gamma\left(k, \delta/2, \Lambda_{\max}, \max_{\Pi \in \Delta}||\Pi^{-1}||\right)\right] \tag{53}$$

where $f_{\text{MM}_k} = f_{\text{MM}_k}\left[\hat{Q}^{2k+1}[Z^n], \hat{\Delta}_l\left(\hat{Q}^{2l+1}[Z^n], \Delta\right)\right]$, the equality is due to (52) and (48), and the inequality is due to (51). On the other hand,

$$P_{\mathbf{Z}}\left(\left|\min_{f \in \mathcal{F}_k} J_k\left(\hat{Q}^{2k+1}[Z^n], (\Delta_n \cup \Delta_{[\eta]}) \cap \mathcal{C}_{\infty}(P_{\mathbf{Z}}), f\right) - \mu_k^{(n)}(P_{\mathbf{Z}}, \Delta_n \cup \Delta_{[\eta]}, \mathbf{Z})\right| > \delta\right)$$
$$\leq \left(|\Delta_n| + |\Delta_{[\eta]}|\right) \cdot \exp\left[-n\Gamma\left(k, \delta, \Lambda_{\max}, \max_{\Pi \in \Delta}||\Pi^{-1}||\right)\right], \tag{54}$$

implying, when combined with (47) and (53) as well as a union bound and the triangle inequality,

$$P_{\mathbf{Z}}\left(\left|\mu_k^{(n)}(P_{\mathbf{Z}}, \Delta_n \cup \Delta_{[\eta]}, \mathbf{Z}) - \mathcal{L}_{\hat{X}_{univ}^n}(P_{\mathbf{Z}}, \Delta_n \cup \Delta_{[\eta]}, \mathbf{Z})\right| > \delta\right)$$
$$= P_{\mathbf{Z}}\left(\left|\mu_k^{(n)}(P_{\mathbf{Z}}, \Delta_n \cup \Delta_{[\eta]}, \mathbf{Z}) - \sup_{\{(P_{\mathbf{X}}, \Pi): \Pi \in \Delta_n \cup \Delta_{[\eta]}, P_{\mathbf{X}} * \Pi = P_{\mathbf{Z}}\}} E_{[P_{\mathbf{X}}, \Pi]}\left[L_{\hat{X}_{\Delta}^{n,k,l}}(X^n, Z^n) | \mathbf{Z}\right]\right| > \delta\right)$$
$$\leq \left(|\Delta_n| + |\Delta_{[\eta]}|\right) \cdot \exp\left[-n\Gamma\left(k, \delta/3, \Lambda_{\max}, \max_{\Pi \in \Delta}||\Pi^{-1}||\right)\right] + U_n(k, l, \eta, \delta/3)$$
$$+ U_n(k, l, \eta, \delta/6) + \left(|\Delta_n| + |\Delta_{[\eta]}|\right) \cdot \exp\left[-n\Gamma\left(k, \delta/6, \Lambda_{\max}, \max_{\Pi \in \Delta}||\Pi^{-1}||\right)\right]. \tag{55}$$

Choosing now $k = k_n$, $l = l_n$, $\eta = \eta_n = \exp(-\sqrt{n}/|\mathcal{A}|^2)$ and noting that $\Delta_{[\eta]}$ can be chosen such that $|\Delta_{[\eta]}| \leq \eta^{-|\mathcal{A}|^2}$ leads to the bound on the right side of (55):

$$2V_n(k_n, l_n, \delta/6) + 2\left(|\Delta_n| + \exp(\sqrt{n})\right) \cdot \exp\left[-n\Gamma\left(k_n, \delta/6, \Lambda_{\max}, \max_{\Pi \in \Delta}||\Pi^{-1}||\right)\right], \tag{56}$$

which is readily verified to be summable for all $\delta > 0$ under the stipulated assumption on the growth rate of $k_n$ and $l_n$.[12] Since $\hat{X}_{univ}^n = \hat{X}_{\Delta}^{n, k_n, l_n}$ we obtain, by the Borel–Cantelli lemma,

$$\lim_{n \to \infty}\left(\mu_{k_n}^{(n)}(P_{\mathbf{Z}}, \Delta_n \cup \Delta_{[\eta_n]}, \mathbf{Z}) - \mathcal{L}_{\hat{X}_{univ}^n}(P_{\mathbf{Z}}, \Delta_n \cup \Delta_{[\eta_n]}, \mathbf{Z})\right) = 0 \qquad P_{\mathbf{Z}} - a.s. \tag{57}$$

---

[12] The growth rate of $k_n$ stipulated in the theorem guarantees that $\exp\left[-n\Gamma\left(k_n, \delta/6, \Lambda_{\max}, \max_{\Pi \in \Delta}||\Pi^{-1}||\right)\right] \leq \exp(-n^{1/2+\varepsilon})$ for an $\varepsilon > 0$ and all sufficiently large $n$. The factor multiplying this exponent $(|\Delta_n| + \exp(\sqrt{n}))$ is upper bounded by $O(\exp(\sqrt{n}))$. Combined with the stipulated summability of $V_n(k_n, l_n, \delta)$ this guarantees the summability of the expression in (56).

Thus we obtain $P_\mathbf{Z}$-a.s.

$$\limsup_{n\to\infty} \left[ \mathcal{L}_{\hat{X}^n_{univ}}(P_\mathbf{Z}, \Delta_n, \mathbf{Z}) - \mu_{k_n}^{(n)}(P_\mathbf{Z}, \Delta, \mathbf{Z}) \right]$$
$$\leq \limsup_{n\to\infty} \left[ \mathcal{L}_{\hat{X}^n_{univ}}(P_\mathbf{Z}, \Delta_n \cup \Delta_{[\eta_n]}, \mathbf{Z}) - \mu_{k_n}^{(n)}(P_\mathbf{Z}, \Delta_n \cup \Delta_{[\eta_n]}, \mathbf{Z}) \right]$$
$$= 0,$$

where the inequality is due to the facts that $\Delta_n \subseteq \Delta_n \cup \Delta_{[\eta_n]} \subseteq \Delta$ and that both $\mathcal{L}_{\hat{X}^n_{univ}}(P_\mathbf{Z}, \cdot, \mathbf{Z})$ and $\mu_{k_n}^{(n)}(P_\mathbf{Z}, \cdot, \mathbf{Z})$ are increasing, and the equality follows from (57).

$\square$

## B Proof of Theorem 1

We start with an outline of the proof idea. The assumption that $\Delta \subseteq \mathcal{C}(P_\mathbf{Z})$ is finite, combined with Lemma 3 and the definition of $J_k$ (recall (1)), imply that, for fixed $k$ and large $n$, $J_k\left(\hat{Q}^{2k+1}[Z^n], \Delta, f\right)$ is uniformly a good estimate of $\mathcal{L}_f^{(n)}(P_\mathbf{Z}, \Delta, \mathbf{Z}) = \sup_{\{(P_\mathbf{X}, \Pi): \Pi \in \Delta, P_\mathbf{X} * \Pi = P_\mathbf{Z}\}} E_{[P_\mathbf{X}, \Pi]}\left[L_f(X^n, Z^n) | Z_{-\infty}^\infty\right]$. Thus, the performance of the sliding window denoiser $f$ that minimizes $G_k\left(\hat{Q}^{2k+1}[Z^n], \Delta, f\right)$ is "close" to $\min_{f \in \mathcal{F}_k} \mathcal{L}_f^{(n)}(P_\mathbf{Z}, \Delta, \mathbf{Z}) = \mu_k^{(n)}(P_\mathbf{Z}, \Delta, \mathbf{Z})$. The bounds in the lemmas of the preceding subsection allow us not only to make this line of argumentation precise, but also to find a rate at which $k$ can be increased with $n$, while maintaining the virtue of the conclusion. In the remainder of this subsection we give the rigorous proof.

For any pair $(P_\mathbf{X}, \Pi)$ such that $\Pi \in \Delta$ and $P_\mathbf{X} * \Pi = P_\mathbf{Z}$, it follows from the definition of $\hat{X}_\Delta^{n,k}$ that

$$E_{[P_\mathbf{X}, \Pi]}\left[L_{\hat{X}_\Delta^{n,k}}(X^n, Z^n) | \mathbf{Z}\right] = E_{[P_\mathbf{X}, \Pi]}\left[L_{f_{\mathbb{MM}_k}[\hat{Q}^{2k+1}[Z^n], \Delta]}(X^n, Z^n) | \mathbf{Z}\right] \quad (58)$$

and, therefore,

$$\max_{\{(P_\mathbf{X}, \Pi): \Pi \in \Delta, P_\mathbf{X} * \Pi = P_\mathbf{Z}\}} E_{[P_\mathbf{X}, \Pi]}\left[L_{\hat{X}_\Delta^{n,k}}(X^n, Z^n) | \mathbf{Z}\right] = \max_{\{(P_\mathbf{X}, \Pi): \Pi \in \Delta, P_\mathbf{X} * \Pi = P_\mathbf{Z}\}} E_{[P_\mathbf{X}, \Pi]}\left[L_{f_{\mathbb{MM}_k}[\hat{Q}^{2k+1}[Z^n], \Delta]}(X^n, Z^n) | \mathbf{Z}\right]. \quad (59)$$

On the other hand, the fact that $\Delta \subseteq \mathcal{C}(P_\mathbf{Z})$ implies that for every $f \in \mathcal{F}_k$

$$\max_{\{(P_\mathbf{X}, \Pi): \Pi \in \Delta, P_\mathbf{X} * \Pi = P_\mathbf{Z}\}} G_k\left(\hat{Q}^{2k+1}[Z^n], \Pi, f\right) = \max_{\Pi \in \Delta} G_k\left(\hat{Q}^{2k+1}[Z^n], \Pi, f\right) = J_k\left(\hat{Q}^{2k+1}[Z^n], \Delta, f\right) \quad (60)$$

implying, when combined with Lemma 4, that

$$P_\mathbf{Z}\left(\max_{f \in \mathcal{F}_k}\left|J_k\left(\hat{Q}^{2k+1}[Z^n], \Delta, f\right) - \max_{\{(P_\mathbf{X}, \Pi): \Pi \in \Delta, P_\mathbf{X} * \Pi = P_\mathbf{Z}\}} E_{[P_\mathbf{X}, \Pi]}[L_f(X^n, Z^n) | \mathbf{Z}]\right| > \delta\right)$$
$$\leq |\Delta| \cdot \exp\left[-n\Gamma\left(k, \delta, \Lambda_{\max}, \max_{\Pi \in \Delta} ||\Pi^{-1}||\right)\right]. \quad (61)$$

Since, by the definition of $f_{\mathbb{MM}_k}$, $J_k\left(\hat{Q}^{2k+1}[Z^n], \Delta, f_{\mathbb{MM}_k}[\hat{Q}^{2k+1}[Z^n], \Delta]\right) = \min_{f \in \mathcal{F}_k} J_k\left(\hat{Q}^{2k+1}[Z^n], \Delta, f\right)$, it follows that

$$P_\mathbf{Z}\left(\left|\min_{f \in \mathcal{F}_k} J_k\left(\hat{Q}^{2k+1}[Z^n], \Delta, f\right) - \max_{\{(P_\mathbf{X}, \Pi): \Pi \in \Delta, P_\mathbf{X} * \Pi = P_\mathbf{Z}\}} E_{[P_\mathbf{X}, \Pi]}\left[L_{\hat{X}_\Delta^{n,k}}(X^n, Z^n) | \mathbf{Z}\right]\right| > \delta\right)$$
$$= P_\mathbf{Z}\left(\left|J_k\left(\hat{Q}^{2k+1}[Z^n], \Delta, f_{\mathbb{MM}_k}[\hat{Q}^{2k+1}[Z^n], \Delta]\right) - \max_{\{(P_\mathbf{X}, \Pi): \Pi \in \Delta, P_\mathbf{X} * \Pi = P_\mathbf{Z}\}} E_{[P_\mathbf{X}, \Pi]}\left[L_{f_{\mathbb{MM}_k}[\hat{Q}^{2k+1}[Z^n], \Delta]}(X^n, Z^n) | \mathbf{Z}\right]\right| > \delta\right)$$
$$\leq |\Delta| \cdot \exp\left[-n\Gamma\left(k, \delta, \Lambda_{\max}, \max_{\Pi \in \Delta} ||\Pi^{-1}||\right)\right], \quad (62)$$

where the equality follows from (59) and the inequality from (61). Furthermore, another application of (61) yields

$$P_{\mathbf{Z}}\left(\left|\min_{f \in \mathcal{F}_k} J_k\left(\hat{Q}^{2k+1}[Z^n], \Delta, f\right) - \mu_k^{(n)}(P_{\mathbf{Z}}, \Delta, \mathbf{Z})\right| > \delta\right)$$
$$= P_{\mathbf{Z}}\left(\left|\min_{f \in \mathcal{F}_k} J_k\left(\hat{Q}^{2k+1}[Z^n], \Delta, f\right) - \min_{f \in \mathcal{F}_k} \max_{\{(P_{\mathbf{X}},\Pi):\Pi \in \Delta, P_{\mathbf{X}}*\Pi = P_{\mathbf{Z}}\}} E_{[P_{\mathbf{X}},\Pi]}\left[L_f(X^n, Z^n)|\mathbf{Z}\right]\right| > \delta\right)$$
$$\leq |\Delta| \cdot \exp\left[-n\Gamma\left(k, \delta, \Lambda_{\max}, \max_{\Pi \in \Delta} ||\Pi^{-1}||\right)\right] \quad (63)$$

which when combined with (62), as well as the triangle inequality and a union bound, implies

$$P_{\mathbf{Z}}\left(\left|\mu_k^{(n)}(P_{\mathbf{Z}}, \Delta, \mathbf{Z}) - \max_{\{(P_{\mathbf{X}},\Pi):\Pi \in \Delta, P_{\mathbf{X}}*\Pi = P_{\mathbf{Z}}\}} E_{[P_{\mathbf{X}},\Pi]}\left[L_{\hat{X}_\Delta^{n,k}}(X^n, Z^n)|\mathbf{Z}\right]\right| > \delta\right)$$
$$\leq 2|\Delta| \cdot \exp\left[-n\Gamma\left(k, \frac{\delta}{2}, \Lambda_{\max}, \max_{\Pi \in \Delta} ||\Pi^{-1}||\right)\right]. \quad (64)$$

Now, the bound on the growth of $k_n$ stipulated in the statement of the theorem is readily verified to guarantee that for every $\delta > 0$, $\sum_n \exp\left[-n\Gamma\left(k_n, \frac{\delta}{2}, \Lambda_{\max}, \max_{\Pi \in \Delta} ||\Pi^{-1}||\right)\right] < \infty$.[13] Recalling that $\hat{X}_{univ}^n = \hat{X}_\Delta^{n,k_n}$, this implies via (64) and the Borel–Cantelli lemma that

$$\lim_{n \to \infty} \left|\mu_{k_n}^{(n)}(P_{\mathbf{Z}}, \Delta, \mathbf{Z}) - \sup_{\{(P_{\mathbf{X}},\Pi):\Pi \in \Delta, P_{\mathbf{X}}*\Pi = P_{\mathbf{Z}}\}} E_{[P_{\mathbf{X}},\Pi]}\left[L_{\hat{X}_{univ}^n}(X^n, Z^n)|\mathbf{Z}\right]\right| = 0 \quad P_{\mathbf{Z}} - a.s.$$

From the notation defined in (14), we see this is exactly (17).

$\square$

## C  Proof of Corollary 2

The proof follows the same lines as the proof of Proposition 1 without the added complexity of an infinite $\Delta$ and having to estimate of $\Delta \cap \mathcal{C}_l(\hat{Q}^{2l+1})$. Hence we will omit the proof of Corollary 2.

## D  Proof of Theorem 2

The main idea is to show that the $\psi$-mixing condition of Theorem 2 implies the conditions on $\alpha$ needed in Lemma 7. Once this is shown, it only remains to appeal to Lemma 7 to conclude the proof. To demonstrate that the $\psi$-mixing condition implies the conditions on $\alpha$, we break the $n$-block into sub-blocks which are separated by uniform gaps. By controlling the rate at which both the sub-blocks and gaps grow with $n$, we can guarantee that the content in the gaps essentially does not effect the empirical distribution, while letting these gaps grow with $n$. We then use the $\psi$-mixing condition and the fact that the gap size is growing with $n$ to drive the joint distribution of the sub-blocks to that of the distribution of independent sub-blocks. This then allows us to uniformly bound the rate of convergence of the empirical distribution to that of the true distribution, which is exactly what is needed for a bound on $\alpha$. We can then apply Lemma 7.

---

[13] The stipulated growth condition is readily seen to imply for any $\varepsilon > 0$ $\exp\left[-nA\left(k, \delta, \Lambda_{\max}, ||\Pi^{-1}||\right)\right] \stackrel{\sim}{<} \exp(-c_\delta n^{3/4-\varepsilon})$, $\exp\left[-nB(k, \delta, \Lambda_{\max}, ||\Pi^{-1}||)\right] \stackrel{\sim}{<} \exp(-c_\delta n^{3/4-\varepsilon/2})$ and, consequently, $\exp\left[-n\Gamma\left(k, \delta, \Lambda_{\max}, \max_{\Pi \in \mathcal{K}} ||\Pi^{-1}||\right)\right] \stackrel{\sim}{<} \exp(-c_\delta n^{1/2})$ (recall (24), (31) and (36) for definitions of these quantities).

Fixing $l$ and $\varepsilon > 0$ we begin by showing bounds on
$$P_{\mathbf{Z}}\left(\left\|P_{Z_{-l}^l} - \hat{Q}^{2l+1}[Z^n]\right\| > \varepsilon\right).$$
Using the union bound we have
$$P_{\mathbf{Z}}\left(\left\|P_{Z_{-l}^l} - \hat{Q}^{2l+1}[Z^n]\right\| > \varepsilon\right) \leq \sum_{x^{2l+1} \in \mathcal{A}^{2l+1}} P_{\mathbf{Z}}\left(\left|P_{Z_{-l}^l}(x^{2l+1}) - \hat{Q}^{2l+1}[Z^n](x^{2l+1})\right| > \varepsilon\right). \tag{65}$$
For each $x^{2l+1} \in \mathcal{A}^{2l+1}$
$$\hat{Q}^{2l+1}[Z^n](x^{2l+1}) = \frac{1}{n_l}\sum_{i=1}^{n_l} Y_i(x^{2l+1}),$$
where $Y_i(x^{2l+1})$ is the indicator function on the event $z_{i(2l+1)+1}^{(2l+1)(i+1)} = x^{2l+1}$ and $n_l = \lfloor n/(2l+1) \rfloor$. For the sake of notational simplicity, we will fix $x^{2l+1} \in \mathcal{A}^{2l+1}$ and use $Y_i$ for $Y_i(x^{2l+1})$. Since $\mathbf{Z}$ is $\psi$-mixing with coefficients $\{\psi_i\}$, then $\mathbf{Y}$ is $\psi$-mixing with coefficients $\psi'_i \leq \psi_{i-2l-1}$ for all $i > 2l+1$.

We now define $S_{n_l} = \sum_{i=1}^{n_l} Y_i$. Therefore we have
$$P_{\mathbf{Z}}\left(\left|P_{Z_{-l}^l}(x^{2l+1}) - \hat{Q}^{2l+1}[Z^n](x^{2l+1})\right| > \varepsilon\right) = P_{\mathbf{Z}}\left(\left|P_{Z_{-l}^l}(x^{2l+1}) - \frac{1}{n_l}S_{n_l}\right| > \varepsilon\right).$$
We can further decompose this as
$$P_{\mathbf{Z}}\left(\left|P_{Z_{-l}^l}(x^{2l+1}) - \hat{Q}^{2l+1}[Z^n](x^{2l+1})\right| > \varepsilon\right) \leq P_{\mathbf{Z}}\left(S_{n_l} > n_l\left(P_{Z_{-l}^l}(x^{2l+1}) + \varepsilon\right)\right) + P_{\mathbf{Z}}\left(S_{n_l} < n_l\left(P_{Z_{-l}^l}(x^{2l+1}) - \varepsilon\right)\right).$$
In order to make use of the Chernoff bound, we rewrite the above as
$$P_{\mathbf{Z}}\left(\left|P_{Z_{-l}^l}(x^{2l+1}) - \hat{Q}^{2l+1}[Z^n](x^{2l+1})\right| > \varepsilon\right) \leq P_{\mathbf{Z}}\left(S_{n_l} > n_l\left(P_{Z_{-l}^l}(x^{2l+1}) + \varepsilon\right)\right)$$
$$+ P_{\mathbf{Z}}\left(n_l - S_{n_l} > n_l\left(1 - P_{Z_{-l}^l}(x^{2l+1}) + \varepsilon\right)\right).$$
Using the Chernoff bound we have
$$P_{\mathbf{Z}}\left(\left|P_{Z_{-l}^l}(x^{2l+1}) - \hat{Q}^{2l+1}[Z^n](x^{2l+1})\right| > \varepsilon\right) \leq E\left[e^{tS_{n_l}}\right]e^{-n_l t(p+\varepsilon)} + E\left[e^{t(n_l - S_{n_l})}\right]e^{-n_l t(\bar{p}+\varepsilon)}, \tag{66}$$
where $p = P_{Z_{-l}^l}(x^{2l+1})$, $\bar{p} = 1-p$ and $t > 0$. Choose $r > 2l+1$ and $m \in \mathbb{N}$ large enough such that
$$1 + \psi'_r = 1 + \psi_{r-2l-1} < \min\left\{e^{1/2D(p+\varepsilon/2||p)}, e^{1/2D(\bar{p}+\varepsilon/2||\bar{p})}\right\}$$
where $D(p+\varepsilon||p)$ is the Kullback Leibler distance between Bernoulli$(p+\varepsilon)$ and Bernoulli$(p)$ distributions, and $m > 2(r+1)/\varepsilon$.

We now turn our attention to bounding $S_{n_l}$. Letting
$$N_{n_l} \triangleq \max\{N \in \mathbb{N} : n_l \geq N(m+r)\}$$
we have
$$\begin{aligned}
S_{n_l} &= \sum_{i=1}^{n_l} Y_i \\
&= \sum_{j=0}^{N_{n_l}-1}\left(\sum_{i=1}^{m} Y_{jN_{n_l}+i} + \sum_{j=1}^{r} Y_{jN_{n_l}+(j+1)m+i}\right) + \sum_{i=N_{n_l}(m+r)+1}^{n_l} Y_i \\
&\stackrel{(a)}{\leq} N_{n_l}(r+1) + \sum_{j=0}^{N_{n_l}-1}\sum_{i=1}^{m} Y_{jN_{n_l}+i},
\end{aligned} \tag{67}$$

where $(a)$ comes from the fact that $Y_i \in \{0, 1\}$ and the definition of $N_{n_l}$. Similarly we can derive the bound

$$S_{n_l} \geq \sum_{j=0}^{N_{n_l}-1} \sum_{i=1}^{m} Y_{jN+i}. \tag{68}$$

Combining (66), (67) and (68) we have

$$P_{\mathbf{Z}}\left(\left|P_{Z_{-l}^l}(x^{2l+1}) - \hat{Q}^{2l+1}[Z^n](x^{2l+1})\right| > \varepsilon\right) \tag{69}$$

$$\leq E\left[\prod_{j=0}^{N_{n_l}} e^{t\sum_{i=1}^{m} Y_{jN_{n_l}+i}}\right] e^{tN_{n_l}(r+1)} e^{-n_l t(p+\varepsilon)} + E\left[e^{tn_l} \prod_{j=0}^{N_{n_l}} e^{-t\sum_{i=1}^{m} Y_{jN_{n_l}+i}}\right] e^{-n_l t(\bar{p}+\varepsilon)}.$$

Since $\mathbf{Y}$ is $\psi$-mixing, we know that the Radon–Nykodim derivative of $(Y_1, \ldots, Y_m)$ and $(Y_{m+r}, \ldots, Y_{2m+r})$ with respect to the product of the marginals is less than or equal to $1 + \psi'_r$. Hence (69) gives us

$$P_{\mathbf{Z}}\left(\left|P_{Z_{-l}^l}(x^{2l+1}) - \hat{Q}^{2l+1}[Z^n](x^{2l+1})\right| > \varepsilon\right)$$

$$\leq (1+\psi'_r)^{N_{n_l}} E\left[e^{tS_m}\right]^{N_{n_l}} e^{tN_{n_l}(r+1)} e^{-n_l t(p+\varepsilon)} + (1+\psi'_r)^{N_{n_l}} E\left[e^{t(m-S_m)}\right]^{N_{n_l}} e^{tN_{n_l}(r+1)} e^{-n_l t(\bar{p}+\varepsilon)}.$$

By our choice of $r$ and $m$ we get

$$P_{\mathbf{Z}}\left(\left|P_{Z_{-l}^l}(x^{2l+1}) - \hat{Q}^{2l+1}[Z^n](x^{2l+1})\right| > \varepsilon\right)$$

$$\leq E\left[e^{tS_m}\right]^{N_{n_l}} e^{-tN_{n_l}m(p+\varepsilon/2)} e^{\frac{N_{n_l}}{2}D(p+\varepsilon/2||p)} + E\left[e^{t'(m-S_m)}\right]^{N_{n_l}} e^{-t'N_{n_l}m(\bar{p}+\varepsilon/2)} e^{\frac{N_{n_l}}{2}D(\bar{p}+\varepsilon/2||\bar{p})}. \tag{70}$$

We also know that $E[e^{tS_m}]$ subject to the constraint that $E[S_m] = mp$ and $S_m \in [0, m]$ is maximized when $S_m$ is m with probability $p$ and 0 with probability $\bar{p}$. Hence

$$E[e^{tS_m}] \leq pe^{mpt} + \bar{p}. \tag{71}$$

Similarly

$$E[e^{t(m-S_m)}] \leq \bar{p}e^{tm\bar{p}} + p. \tag{72}$$

Combining (70), (71) and (72) we get

$$P_{\mathbf{Z}}\left(\left|P_{Z_{-l}^l}(x^{2l+1}) - \hat{Q}^{2l+1}[Z^n](x^{2l+1})\right| > \varepsilon\right)$$

$$\leq \left[(pe^{mpt} + \bar{p}) e^{-tm(p+\varepsilon/2)}\right]^{N_{n_l}} e^{\frac{N_{n_l}}{2}D(p+\varepsilon/2||p)} + \left[\left(\bar{p}e^{m\bar{p}t'} + p\right) e^{-t'm(\bar{p}+\varepsilon/2)}\right]^{N_{n_l}} e^{\frac{N_{n_l}}{2}D(\bar{p}+\varepsilon/2||\bar{p})}.$$

Since the above equation is true for all $t, t' > 0$ we can take the infimum over all $t, t' > 0$ and get

$$P_{\mathbf{Z}}\left(\left|P_{Z_{-l}^l}(x^{2l+1}) - \hat{Q}^{2l+1}[Z^n](x^{2l+1})\right| > \varepsilon\right) \tag{73}$$

$$\leq e^{\frac{N_{n_l}}{2}D(p+\varepsilon/2||p)} \left[\inf_{t>0} \left(pe^{mpt} + \bar{p}\right) e^{-tm(p+\varepsilon/2)}\right]^{N_{n_l}} + e^{\frac{N_{n_l}}{2}D(\bar{p}+\varepsilon/2||\bar{p})} \left[\inf_{t'>0} \left(\bar{p}e^{m\bar{p}t'} + p\right) e^{-t'm(\bar{p}+\varepsilon/2)}\right]^{N_{n_l}}.$$

Since $D(p+\varepsilon||p)$ is the rate function for a Bernoulli($p$) process, it follows that the infimum in (73) yields

$$P_{\mathbf{Z}}\left(\left|P_{Z_{-l}^l}(x^{2l+1}) - \hat{Q}^{2l+1}[Z^n](x^{2l+1})\right| > \varepsilon\right) \leq e^{-\frac{N_{n_l}}{2}D(p+\varepsilon/2||p)} + e^{-\frac{N_{n_l}}{2}D(\bar{p}+\varepsilon/2||\bar{p})}. \tag{74}$$

We can now further upper bound by taking the maximum over $p$. Letting

$$p_1^*(a) = \arg\min_{p \in [0,1]} D(p+a||p),$$

further bounding of (74) yields

$$P_{\mathbf{Z}}\left(\left|P_{Z_{-l}^l}(x^{2l+1}) - \hat{Q}^{2l+1}[Z^n](x^{2l+1})\right| > \varepsilon\right) \leq 2e^{-\frac{N_{n_l}}{2}D(p^*(\varepsilon/2)+\varepsilon/2||p^*(\varepsilon/2))}. \tag{75}$$

Since (75) is true for all $x^{2l+1} \in \mathcal{A}^{2l+1}$, then (65), (75), and the definition of $N_{n_l}$ yield

$$P_{\mathbf{Z}}\left(\left\|P_{Z_{-l}^l} - \hat{Q}^{2l+1}[Z^n]\right\| > \varepsilon\right) \leq 2|\mathcal{A}|^{2l+1} e^{-\frac{1}{2}\left(\frac{n}{(m+r)(2l+1)}-2\right)D(p^*(\varepsilon/2)+\varepsilon/2||p^*(\varepsilon/2))}. \tag{76}$$

Further upper bounding $D(p^*(\varepsilon/2) + \varepsilon/2||p^*(\varepsilon/2))$ by $D(1/2+\varepsilon||1/2) < \log(1+2\varepsilon)$ we obtain

$$P_{\mathbf{Z}}\left(\left\|P_{Z_{-l}^l} - \hat{Q}^{2l+1}[Z^n]\right\| > \varepsilon\right) \leq 2(1+2\varepsilon)|\mathcal{A}|^{2l+1} e^{-\frac{1}{2}\frac{n}{(m+r)(2l+1)}D(p^*(\varepsilon/2)+\varepsilon/2||p^*(\varepsilon/2))}, \tag{77}$$

the bound in (77) being valid for all $n$ (since if $n < (m+r)(2l+1)$ the bound is greater than 1). Note without loss of generality assume $\varepsilon \leq 1$. Hence $P_{\mathbf{Z}} \in \tilde{\mathcal{S}}(\mathcal{A}, \alpha_\psi)$ where $\alpha_\psi$ is defined by

$$\alpha_\psi(n,l,\varepsilon) = 6|\mathcal{A}|^{2l+1} e^{-\frac{1}{2}\frac{n}{(m+r)(2l+1)}D(p^*(\varepsilon/2)+\varepsilon/2||p^*(\varepsilon/2))} \tag{78}$$

with $r > 2l+1$ and $m \in \mathbb{N}$ chosen such that

$$1 + \psi_r' = 1 + \psi_{r-2l-1} < e^{1/2 D(p^*(\varepsilon/2)+\varepsilon/2||p^*(\varepsilon/2))},$$

and $m > 2(r+1)/\varepsilon$.

We now turn to bounding $V_n$ as defined in (39). We first define the following

$$\begin{aligned}
C_k^{(1)} &= |\mathcal{A}|^{2k+1} \Lambda_{\max} \max_{\Pi \in \Delta} ||\Pi^{-1}|| \\
C_l^{(2)} &= \left(\max_{\Pi \in \Delta} ||\Pi^{-1}||\right)^{-|\mathcal{A}|^l} \\
\eta &= e^{-\frac{\sqrt{n}}{|\mathcal{A}|^2}}.
\end{aligned}$$

For $\delta > 0$, we can now expand $V_n$ as follows

$$\begin{aligned}
V_n(k,l,\delta) &= \alpha_\psi\left(n, l, b_l^{-1}\left(\phi_k^{-1}(\delta - \phi_k(\eta)) - a_l\right)\right) \\
&= \alpha_\psi\left(n, l, \frac{C_l^{(2)}}{C_k^{(1)}}\delta - C_l^{(2)}\eta - C_l^{(1)} a_l\right)
\end{aligned} \tag{79}$$

For a given sequence $\{l_n\}$, choose

$$k_n \leq \min\left\{\frac{\ln n}{16 \ln |\mathcal{A}|}, -\frac{\ln a_{l_n}}{4 \ln |\mathcal{A}|}\right\}.$$

This restriction on $k_n$ assures us that there exits $N'$ such that

$$\frac{C_{l_n}^{(2)}}{C_{k_n}^{(1)}}\delta - C_{l_n}^{(2)}\eta - C_{l_n}^{(1)} a_{l_n} > 0 \quad \forall n \geq N'.$$

We now choose
$$g_n \geq \max\{4n^{1/3}, -\ln a_{l_n}, l_n\}$$
and define
$$\varepsilon_n = |\mathcal{A}|^{-g_n}.$$
Notice that $\varepsilon_n$ is monotonically decreasing to 0 and that $\varepsilon_n$ is independent of $\delta$. Also, by our choice of $g_n$, we are assured that there exists $N''$ such that
$$\frac{C_{l_n}^{(2)}}{C_{k_n}^{(1)}}\delta - C_{l_n}^{(2)}\eta - C_{l_n}^{(1)}a_{l_n} > \varepsilon_n \quad \forall n \geq N''. \tag{80}$$

Combining the monotonicity of $\alpha_\psi(n, l, \varepsilon)$ in $\varepsilon$ and (80) gives the following: If
$$\sum_{i=1}^{\infty} \alpha_\psi(n, l_n, \varepsilon_n) < \infty, \tag{81}$$
then
$$\sum_{i=1}^{\infty} V_n(l_n, l_n, \delta) < \infty \quad \forall \, \delta > 0.$$

We now construct an unbounded sequence $\{w_n\}_{n=1}^{\infty}$. For $n$ small, $w_n$ can be chosen arbitrary. For $n$ large, let $w_n$ be defined such that
$$(2w_n + 1)\ln|\mathcal{A}| - \frac{n}{2}C(w_n, \{\psi_i\}_{i=1}^{\infty}) < -2 \tag{82}$$
where
$$C(w_n, \{\psi_i\}_{i=1}^{\infty}) = \frac{D\left(p^*\left(\varepsilon_{w_n}/2\right) + \varepsilon_{w_n}/2 \,\middle|\middle|\, p^*\left(\varepsilon_{w_n}/2\right)\right)}{(m_{w_n} + r_{w_n})(2w_n + 1)},$$
with $m_{w_n}, r_{w_n} \in \{1, 2, \ldots, n\}$ chosen such that $r_{w_n} > 2w_n + 1$,
$$1 + \psi_{r_{w_n} - 2w_n - 1} < e^{1/2 D(p^*(\varepsilon_{w_n}/2) + \varepsilon_{w_n}/2 \,||\, p^*(\varepsilon_{w_n}/2))}, \tag{83}$$
and
$$m_{w_n} > \frac{2(r_{w_n} + 1)}{\varepsilon_{w_n}}.$$

Notice that both $(2w_n + 1)\ln|\mathcal{A}|$ and $C(w_n, \{\psi_i\}_{i=1}^{\infty})$ are decreasing in $w_n$. Furthermore, their dependence on $n$ comes only through the sequence $\{w_n\}$. Hence combining the fact that $\psi_r \to 0$ and by allowing $w_n$ to grow slowly with respect to $n$, we can insure that inequality (82) holds.

Expanding $\alpha_\psi(n, l_n, \varepsilon_n)$, we see that (81) holds whenever $\{l_n\}$ and $\{k_n\}$ are unbounded sequences such that
$$l_n \leq w_n \tag{84}$$
and
$$k_n \leq \min\left\{\frac{\ln n}{16 \ln|\mathcal{A}|}, -\frac{\ln a_{l_n}}{4 \ln|\mathcal{A}|}\right\}.$$

Note, since $\{w_n\}$ is unbounded and from Lemma 6 we know that $a_l \to 0$, we can choose $\{l_n\}$ and $\{k_n\}$ to be unbounded. Recall that $a_l$ is used to denote $a_l(P_{\mathbf{Z}}, \Delta)$ and is a function of the distribution $P_{\mathbf{Z}}$. Hence although the constraint on $\{l_n\}$ is independent of $P_{\mathbf{Z}}$, the constraint on $\{k_n\}$ is not. However, from Lemma 6 we know that $a_l(P_{\mathbf{Z}}, \Delta) \to 0$ uniformly in $P_{\mathbf{Z}}$. Uniform convergence implies

$$\lim_{l \to \infty} \sup_{P_{\mathbf{Z}} \in \tilde{\mathcal{S}}} a_l(P_{\mathbf{Z}}, \Delta) = 0.$$

We can therefore choose $\{k_n\}$ independent of $\{a_l(P_{\mathbf{Z}}, \Delta)\}$ and hence independent of $P_{\mathbf{Z}}$. In particular, we can choose $\{k_n\}$ unbounded and satisfying

$$k_n \leq \min \left\{ \frac{\ln n}{16 \ln |\mathcal{A}|}, -\sup_{P_{\mathbf{Z}} \in \tilde{\mathcal{S}}} \frac{\ln a_{l_n}(P_{\mathbf{Z}}, \Delta)}{4 \ln |\mathcal{A}|} \right\}. \tag{85}$$

Theorem 2 now follows by applying Lemma 7 for any unbounded sequences $\{k_n\}$ and $\{l_n\}$ satisfying (84) and (85).

$\square$

## E Proof of Proposition 1

The idea of the proof that follows is to combine Lemma 1, Lemma 2, and the triangle inequality to get a bound on the terms of the limit in (23), and then to use Lemma 7 to show that the bound vanishes in the limit.

Before going through the proof, we note that by the same argument as in the proof of Theorem 2, we can construct sequences $\{k_n\}$ and $\{l_n\}$ such that for all $P_{\mathbf{Z}} \in \tilde{\mathcal{S}}(\mathcal{A}, \alpha_\psi)$,

$$\sum_{n=1}^{\infty} \alpha_\psi \left( n, k_n, \frac{\varepsilon}{\max_{\Pi \in \Delta} ||\Pi^{-1}|| \Lambda_{\max} |\mathcal{A}|^{6k_n + 3}} \right) \tag{86}$$

$$\leq \sum_{n=1}^{\infty} \alpha_\psi (n, k_n, \varepsilon_n) < \infty \quad \forall \varepsilon > 0, \tag{87}$$

where $\varepsilon_n$ and $\alpha_\psi$ are defined as in the proof of Theorem 2.

Lemma 7 gives us

$$\lim_{n \to \infty} \left[ \mathcal{L}_{\hat{X}^n_{univ}} (P_{\mathbf{Z}}, \Delta, \mathbf{Z}) - \mu^{(n)}_{k_n} (P_{\mathbf{Z}}, \Delta, \mathbf{Z}) \right] = 0 \qquad P_{\mathbf{Z}} - a.s.$$

Letting $E_{\mathbf{Z}}$ stand for expectation under $P_{\mathbf{Z}}$ and taking expectation in the above equality, it follows from the bounded convergence theorem that

$$\lim_{n \to \infty} \left[ E_{\mathbf{Z}} \left[ \mathcal{L}_{\hat{X}^n_{univ}} (P_{\mathbf{Z}}, \Delta, \mathbf{Z}) \right] - E_{\mathbf{Z}} \left[ \mu^{(n)}_{k_n} (P_{\mathbf{Z}}, \Delta, \mathbf{Z}) \right] \right] = 0.$$

Expanding the inner terms gives

$$\lim_{n \to \infty} \left[ E_{\mathbf{Z}} \left[ \max_{(P_{\mathbf{X}}, \Pi)} E_{[P_{\mathbf{X}}, \Pi]} \left[ L_{\hat{X}^n_{univ}} (X^n, Z^n) | \mathbf{Z} \right] \right] - E_{\mathbf{Z}} \left[ \min_{f \in \mathcal{F}_{k_n}} \max_{(P_{\mathbf{X}}, \Pi)} E_{[P_{\mathbf{X}}, \Pi]} \left[ L_f (X^n, Z^n) | \mathbf{Z} \right] \right] \right] = 0,$$

where for notational simplicity, we suppress the constraints on $(P_{\mathbf{X}}, \Pi)$ in the maximization. Moving the expectations in we get

$$\limsup_{n \to \infty} \left[ \max_{(P_{\mathbf{X}}, \Pi)} E_{[P_{\mathbf{X}}, \Pi]} \left[ L_{\hat{X}^n_{univ}} (X^n, Z^n) \right] - \min_{f \in \mathcal{F}_{k_n}} E_{\mathbf{Z}} \left[ \max_{(P_{\mathbf{X}}, \Pi)} E_{[P_{\mathbf{X}}, \Pi]} \left[ L_f (X^n, Z^n) | \mathbf{Z} \right] \right] \right] \leq 0. \tag{88}$$

Defining
$$B_n = \min_{f \in \mathcal{F}_{k_n}} \max_{(P_{\mathbf{X}}, \Pi)} E_{[P_{\mathbf{X}}, \Pi]} \left[ L_f(X^n, Z^n) \right] - \min_{f \in \mathcal{F}_{k_n}} E_{\mathbf{Z}} \left[ \max_{(P_{\mathbf{X}}, \Pi)} E_{[P_{\mathbf{X}}, \Pi]} \left[ L_f(X^n, Z^n) | \mathbf{Z} \right] \right],$$

(88) gives us

$$\limsup_{n \to \infty} \left[ \max_{(P_{\mathbf{X}}, \Pi)} E_{[P_{\mathbf{X}}, \Pi]} \left[ L_{\hat{X}^n_{univ}}(X^n, Z^n) \right] - \min_{f \in \mathcal{F}_{k_n}} \max_{(P_{\mathbf{X}}, \Pi)} E_{[P_{\mathbf{X}}, \Pi]} \left[ L_f(X^n, Z^n) \right] + B_n \right] \leq 0. \tag{89}$$

For notational convenience denote

$$g_{P_{\mathbf{X}}, \Pi, f}(\mathbf{Z}) \triangleq E_{[P_{\mathbf{X}}, \Pi]} \left[ L_f(X^n, Z^n) | \mathbf{Z} \right].$$

Let $\delta > 0$

$$\left| E_{\mathbf{Z}} [g_{P_{\mathbf{X}}, \Pi, f}(\mathbf{Z})] - G_{k_n} \left( \hat{Q}^{2k_n+1}[Z^n], \Pi, f \right) \right| = \left| E_{[P_{\mathbf{X}}, \Pi]} [L_f(X^n, Z^n)] - G_{k_n} \left( \hat{Q}^{2k_n+1}[Z^n], \Pi, f \right) \right|$$
$$\leq \left| E_{[P_{\mathbf{X}}, \Pi]} \left[ L_f(X^n, Z^n) - G_{k_n} \left( \hat{Q}^{2k_n+1}[Z^n], \Pi, f \right) \right] \right|$$
$$+ \left| E_{[P_{\mathbf{X}}, \Pi]} \left[ G_{k_n} \left( \hat{Q}^{2k_n+1}[Z^n], \Pi, f \right) \right] - G_{k_n} \left( \hat{Q}^{2k_n+1}[Z^n], \Pi, f \right) \right|$$
$$\leq E_{[P_{\mathbf{X}}, \Pi]} \left[ \left| L_f(X^n, Z^n) - G_{k_n} \left( \hat{Q}^{2k_n+1}[Z^n], \Pi, f \right) \right| \right]$$
$$+ \left| E_{[P_{\mathbf{X}}, \Pi]} \left[ G_{k_n} \left( \hat{Q}^{2k_n+1}[Z^n], \Pi, f \right) \right] - G_{k_n} \left( \hat{Q}^{2k_n+1}[Z^n], \Pi, f \right) \right|$$

$$\left| E_{\mathbf{Z}} [g_{P_{\mathbf{X}}, \Pi, f}(\mathbf{Z})] - G_{k_n} \left( \hat{Q}^{2k_n+1}[Z^n], \Pi, f \right) \right| \overset{(a)}{\leq} \Lambda_{\max} e^{-n A(k_n, \delta, \Lambda_{\max}, ||\Pi^{-1}||)} + \delta \tag{90}$$
$$+ \left| E_{[P_{\mathbf{X}}, \Pi]} \left[ G_{k_n} \left( \hat{Q}^{2k_n+1}[Z^n], \Pi, f \right) \right] - G_{k_n} \left( \hat{Q}^{2k_n+1}[Z^n], \Pi, f \right) \right|$$

where $(a)$ follows from lemma 1. Since (90) holds for all $(P_{\mathbf{X}}, \Pi, f)$ we have

$$\max_{f \in \mathcal{F}_{k_n}} \max_{(P_{\mathbf{X}}, \Pi)} \left| E_{\mathbf{Z}} [g_{P_{\mathbf{X}}, \Pi, f}(\mathbf{Z})] - G_{k_n} \left( \hat{Q}^{2k_n+1}[Z^n], \Pi, f \right) \right| \leq$$
$$\Lambda_{\max} e^{-n A(k_n, \delta, \Lambda_{\max}, \max_{\Pi \in \Delta} ||\Pi^{-1}||)} + \delta + \max_{f \in \mathcal{F}_{k_n}} \max_{(P_{\mathbf{X}}, \Pi)} \left| E_{[P_{\mathbf{X}}, \Pi]} \left[ G_{k_n} \left( \hat{Q}^{2k_n+1}[Z^n], \Pi, f \right) \right] - G_{k_n} \left( \hat{Q}^{2k_n+1}[Z^n], \Pi, f \right) \right|.$$
(91)

To proceed, we establish the following.

**Claim 1**

$$\limsup_{n \to \infty} \max_{f \in \mathcal{F}_{k_n}} \max_{(P_{\mathbf{X}}, \Pi)} \left| E_{[P_{\mathbf{X}}, \Pi]} \left[ G_{k_n} \left( \hat{Q}^{2k_n+1}[Z^n], \Pi, f \right) \right] - G_{k_n} \left( \hat{Q}^{2k_n+1}[Z^n], \Pi, f \right) \right| = 0 \quad P_{\mathbf{Z}} - a.s.$$

*Proof of Claim 1:*

The definition of $G_k$ is readily seen to imply

$$\left| E_{[P_{\mathbf{X}}, \Pi]} \left[ G_{k_n} \left( \hat{Q}^{2k_n+1}[Z^n], \Pi, f \right) \right] - G_{k_n} \left( \hat{Q}^{2k_n+1}[Z^n], \Pi, f \right) \right| \leq$$
$$\max_{\Pi \in \Delta} ||\Pi^{-1}|| \Lambda_{\max} |\mathcal{A}|^{6k_n+3} \left| \left| E_{\mathbf{Z}} \left[ \hat{Q}^{2k_n+1}[Z^n] \right] - \hat{Q}^{2k_n+1}[Z^n] \right| \right| =$$
$$\max_{\Pi \in \Delta} ||\Pi^{-1}|| \Lambda_{\max} |\mathcal{A}|^{6k_n+3} \left| \left| P_{Z^{k_n}_{-k_n}} - \hat{Q}^{2k_n+1}[Z^n] \right| \right|.$$

By the construction of $\alpha_\psi$, for any $\varepsilon > 0$ we have

$$P_{\mathbf{Z}}\left(\left|E_{[P_{\mathbf{X}},\Pi]}\left[G_{k_n}\left(\hat{Q}^{2k_n+1}[Z^n],\Pi,f\right)\right] - G_{k_n}\left(\hat{Q}^{2k_n+1}[Z^n],\Pi,f\right)\right| > \varepsilon\right) \leq \alpha_\psi\left(n, k_n, \frac{\varepsilon}{\max_{\Pi \in \Delta} ||\Pi^{-1}||\Lambda_{\max}|\mathcal{A}|^{6k_n+3}}\right).$$

Since by hypothesis the right hand side is summable, the Borel–Cantelli lemma implies

$$\lim_{n \to \infty}\left|E_{[P_{\mathbf{X}},\Pi]}\left[G_{k_n}\left(\hat{Q}^{2k_n+1}[Z^n],\Pi,f\right)\right] - G_{k_n}\left(\hat{Q}^{2k_n+1}[Z^n],\Pi,f\right)\right| = 0 \quad P_{\mathbf{Z}} - a.s.$$

Note that for each $n$,

$$\left|E_{[P_{\mathbf{X}},\Pi]}\left[G_{k_n}\left(\hat{Q}^{2k_n+1}[Z^n],\Pi,f\right)\right] - G_{k_n}\left(\hat{Q}^{2k_n+1}[Z^n],\Pi,f\right)\right|$$

is continuous in $f$, that $f \in \mathcal{F}_{k_n} \subset \mathcal{F}_\infty$, and by Tychonoff's Theorem $\mathcal{F}_\infty$ is compact. We can therefore apply Dini's Theorem which implies that the limit is uniform in $f$. Due to the finiteness of $|\Delta|$, uniform convergence in $f$ implies the convergence is uniform in $(P_{\mathbf{X}}, \Pi, f)$, thus establishing the Claim. □

Returning to the proof, the combination of Claim 1 and (91) gives

$$\limsup_{n \to \infty} \max_{f \in \mathcal{F}_{k_n}} \max_{(P_{\mathbf{X}},\Pi)} \left|E_{\mathbf{Z}}[g_{P_{\mathbf{X}},\Pi,f}(\mathbf{Z})] - G_{k_n}\left(\hat{Q}^{2k_n+1}[Z^n],\Pi,f\right)\right| \leq$$

$$\limsup_{n \to \infty} \Lambda_{\max} e^{-nA(k_n,\delta,\Lambda_{\max},\max_{\Pi \in \Delta}||\Pi^{-1}||)} + \delta \quad P_{\mathbf{Z}} - a.s.$$

Since $k_n$ is chosen such that $k_n \leq \frac{\ln(n)}{16 \ln |\mathcal{A}|}$, $e^{-nA(k_n,\delta,\Lambda_{\max},\max_{\Pi \in \Delta}||\Pi^{-1}||)} \to 0$ (recall lemma 1 for what $A$ can be chosen to be) and therefore

$$\limsup_{n \to \infty} \max_{f \in \mathcal{F}_{k_n}} \max_{(P_{\mathbf{X}},\Pi)} \left|E_{\mathbf{Z}}[g_{P_{\mathbf{X}},\Pi,f}(\mathbf{Z})] - G_{k_n}\left(\hat{Q}^{2k_n+1}[Z^n],\Pi,f\right)\right| \leq \delta \quad P_{\mathbf{Z}} - a.s.$$

implying

$$\lim_{n \to \infty} \max_{f \in \mathcal{F}_{k_n}} \max_{(P_{\mathbf{X}},\Pi)} \left|E_{\mathbf{Z}}[g_{P_{\mathbf{X}},\Pi,f}(\mathbf{Z})] - G_{k_n}\left(\hat{Q}^{2k_n+1}[Z^n],\Pi,f\right)\right| = 0 \quad P_{\mathbf{Z}} - a.s.$$

by the arbitrariness of $\delta$. We also note that

$$\left|\min_{f \in \mathcal{F}_{k_n}} \max_{(P_{\mathbf{X}},\Pi)} E_{\mathbf{Z}}[g_{P_{\mathbf{X}},\Pi,f}(\mathbf{Z})] - \min_{f \in \mathcal{F}_{k_n}} \max_{(P_{\mathbf{X}},\Pi)} G_{k_n}\left(\hat{Q}^{2k_n+1}[Z^n],\Pi,f\right)\right| \leq$$

$$\max_{f \in \mathcal{F}_{k_n}} \max_{(P_{\mathbf{X}},\Pi)} \left|E_{\mathbf{Z}}[g_{P_{\mathbf{X}},\Pi,f}(\mathbf{Z})] - G_{k_n}\left(\hat{Q}^{2k_n+1}[Z^n],\Pi,f\right)\right|$$

and therefore

$$\lim_{n \to \infty} \left|\min_{f \in \mathcal{F}_{k_n}} \max_{(P_{\mathbf{X}},\Pi)} E_{\mathbf{Z}}[g_{P_{\mathbf{X}},\Pi,f}(\mathbf{Z})] - \min_{f \in \mathcal{F}_{k_n}} \max_{(P_{\mathbf{X}},\Pi)} G_{k_n}\left(\hat{Q}^{2k_n+1}[Z^n],\Pi,f\right)\right| = 0 \quad P_{\mathbf{Z}} - a.s. \quad (92)$$

From lemma 2 we have

$$P_{\mathbf{Z}}\left(\left|G_{k_n}\left(\hat{Q}^{2k_n+1}[Z^n],\Pi,f\right) - g_{P_{\mathbf{X}},\Pi,f}(\mathbf{Z})\right| > \delta\right) \leq e^{-nB(k_n,\delta,\Lambda_{\max},||\Pi^{-1}||)}.$$

Therefore, applying the union bound, we obtain

$$P_{\mathbf{Z}}\left(\max_{(P_{\mathbf{X}},\Pi)} \left|G_{k_n}\left(\hat{Q}^{2k_n+1}[Z^n],\Pi,f\right) - g_{P_{\mathbf{X}},\Pi,f}(\mathbf{Z})\right| > \delta\right) \leq |\Delta| e^{-nB(k_n,\delta,\Lambda_{\max},\max_{\Pi \in \Delta}||\Pi^{-1}||)}.$$

Since
$$\max_{(P_{\mathbf{X}},\Pi)} \left|G_{k_n}\left(\hat{Q}^{2k_n+1}[Z^n],\Pi,f\right) - g_{P_{\mathbf{X}},\Pi,f}(\mathbf{Z})\right| \geq \left|\max_{(P_{\mathbf{X}},\Pi)} G_{k_n}\left(\hat{Q}^{2k_n+1}[Z^n],\Pi,f\right) - \max_{(P_{\mathbf{X}},\Pi)} g_{P_{\mathbf{X}},\Pi,f}(\mathbf{Z})\right|$$

we have
$$P_{\mathbf{Z}}\left(\left|\max_{(P_{\mathbf{X}},\Pi)} G_{k_n}\left(\hat{Q}^{2k_n+1}[Z^n],\Pi,f\right) - \max_{(P_{\mathbf{X}},\Pi)} g_{P_{\mathbf{X}},\Pi,f}(\mathbf{Z})\right| > \delta\right) \leq |\Delta|e^{-nB(k_n,\delta,\Lambda_{\max},\max_{\Pi\in\Delta}||\Pi^{-1}||)}.$$

Hence
$$\left|E_{\mathbf{Z}}\left[\max_{(P_{\mathbf{X}},\Pi)} G_{k_n}\left(\hat{Q}^{2k_n+1}[Z^n],\Pi,f\right)\right] - E_{\mathbf{Z}}\left[\max_{(P_{\mathbf{X}},\Pi)} g_{P_{\mathbf{X}},\Pi,f}(\mathbf{Z})\right]\right| \leq \Lambda_{\max}|\Delta|e^{-nB(k_n,\delta,\Lambda_{\max},\max_{\Pi\in\Delta}||\Pi^{-1}||)} + \delta.$$

Since this is true for all $f \in \mathcal{F}_{k_n}$ we have
$$\max_{f\in\mathcal{F}_{k_n}}\left|E_{\mathbf{Z}}\left[\max_{(P_{\mathbf{X}},\Pi)} G_{k_n}\left(\hat{Q}^{2k_n+1}[Z^n],\Pi,f\right)\right] - E_{\mathbf{Z}}\left[\max_{(P_{\mathbf{X}},\Pi)} g_{P_{\mathbf{X}},\Pi,f}(\mathbf{Z})\right]\right| \leq \Lambda_{\max}|\Delta|e^{-nB(k_n,\delta,\Lambda_{\max},\max_{\Pi\in\Delta}||\Pi^{-1}||)}+\delta.$$

Since
$$\max_{f\in\mathcal{F}_{k_n}}\left|E_{\mathbf{Z}}\left[\max_{(P_{\mathbf{X}},\Pi)} G_{k_n}\left(\hat{Q}^{2k_n+1}[Z^n],\Pi,f\right)\right] - E_{\mathbf{Z}}\left[\max_{(P_{\mathbf{X}},\Pi)} g_{P_{\mathbf{X}},\Pi,f}(\mathbf{Z})\right]\right| \geq$$
$$\left|\min_{f\in\mathcal{F}_{k_n}} E_{\mathbf{Z}}\left[\max_{(P_{\mathbf{X}},\Pi)} G_{k_n}\left(\hat{Q}^{2k_n+1}[Z^n],\Pi,f\right)\right] - \min_{f\in\mathcal{F}_{k_n}} E_{\mathbf{Z}}\left[\max_{(P_{\mathbf{X}},\Pi)} g_{P_{\mathbf{X}},\Pi,f}(\mathbf{Z})\right]\right|,$$

we also have
$$\left|\min_{f\in\mathcal{F}_{k_n}} E_{\mathbf{Z}}\left[\max_{(P_{\mathbf{X}},\Pi)} G_{k_n}\left(\hat{Q}^{2k_n+1}[Z^n],\Pi,f\right)\right] - \min_{f\in\mathcal{F}_{k_n}} E_{\mathbf{Z}}\left[\max_{(P_{\mathbf{X}},\Pi)} g_{P_{\mathbf{X}},\Pi,f}(\mathbf{Z})\right]\right| \leq$$
$$\Lambda_{\max}|\Delta|e^{-nB(k_n,\delta,\Lambda_{\max},\max_{\Pi\in\Delta}||\Pi^{-1}||)} + \delta$$

and therefore
$$\limsup_{n\to\infty}\left|\min_{f\in\mathcal{F}_{k_n}} E_{\mathbf{Z}}\left[\max_{(P_{\mathbf{X}},\Pi)} G_{k_n}\left(\hat{Q}^{2k_n+1}[Z^n],\Pi,f\right)\right] - \min_{f\in\mathcal{F}_{k_n}} E_{\mathbf{Z}}\left[\max_{(P_{\mathbf{X}},\Pi)} g_{P_{\mathbf{X}},\Pi,f}(\mathbf{Z})\right]\right| \leq$$
$$\limsup_{n\to\infty} \Lambda_{\max}|\Delta|e^{-nB(k_n,\delta,\Lambda_{\max},\max_{\Pi\in\Delta}||\Pi^{-1}||)} + \delta$$

Since $k_n$ is chosen such that $k_n \leq \frac{\ln(n)}{16\ln|\mathcal{A}|}$, lemma 2 implies that $B$ can be chosen such that $e^{-nB(k_n,\delta,\Lambda_{\max},\max_{\Pi\in\Delta}||\Pi^{-1}||)} \to 0$ and therefore
$$\limsup_{n\to\infty}\left|\min_{f\in\mathcal{F}_{k_n}} E_{\mathbf{Z}}\left[\max_{(P_{\mathbf{X}},\Pi)} G_{k_n}\left(\hat{Q}^{2k_n+1}[Z^n],\Pi,f\right)\right] - \min_{f\in\mathcal{F}_{k_n}} E_{\mathbf{Z}}\left[\max_{(P_{\mathbf{X}},\Pi)} g_{P_{\mathbf{X}},\Pi,f}(\mathbf{Z})\right]\right| \leq \delta,$$

implying, by the arbitrariness of $\delta > 0$,
$$\lim_{n\to\infty}\left|\min_{f\in\mathcal{F}_{k_n}} E_{\mathbf{Z}}\left[\max_{(P_{\mathbf{X}},\Pi)} G_{k_n}\left(\hat{Q}^{2k_n+1}[Z^n],\Pi,f\right)\right] - \min_{f\in\mathcal{F}_{k_n}} E_{\mathbf{Z}}\left[\max_{(P_{\mathbf{X}},\Pi)} g_{P_{\mathbf{X}},\Pi,f}(\mathbf{Z})\right]\right| = 0. \quad (93)$$

Before completing the proof, we shall need to establish the following.

**Claim 2**

$$\lim_{n\to\infty}\left|\min_{f\in\mathcal{F}_{k_n}} E_{\mathbf{Z}}\left[\max_{(P_{\mathbf{X}},\Pi)} G_{k_n}\left(\hat{Q}^{2k_n+1}[Z^n],\Pi,f\right)\right] - \min_{f\in\mathcal{F}_{k_n}} \max_{(P_{\mathbf{X}},\Pi)} G_{k_n}\left(\hat{Q}^{2k_n+1}[Z^n],\Pi,f\right)\right| = 0 \quad P_{\mathbf{Z}} - a.s.$$

*Proof of Claim 2:*

Since

$$\left|\min_{f\in\mathcal{F}_{k_n}} E_{\mathbf{Z}}\left[\max_{(P_{\mathbf{X}},\Pi)} G_{k_n}\left(\hat{Q}^{2k_n+1}[Z^n],\Pi,f\right)\right] - \min_{f\in\mathcal{F}_{k_n}}\max_{(P_{\mathbf{X}},\Pi)} G_{k_n}\left(\hat{Q}^{2k_n+1}[Z^n],\Pi,f\right)\right| \leq$$

$$\max_{f\in\mathcal{F}_{k_n}}\left|E_{\mathbf{Z}}\left[\max_{(P_{\mathbf{X}},\Pi)} G_{k_n}\left(\hat{Q}^{2k_n+1}[Z^n],\Pi,f\right)\right] - \max_{(P_{\mathbf{X}},\Pi)} G_{k_n}\left(\hat{Q}^{2k_n+1}[Z^n],\Pi,f\right)\right|,$$

it is sufficient to show

$$\lim_{n\to\infty}\max_{f\in\mathcal{F}_{k_n}}\left|E_{\mathbf{Z}}\left[\max_{(P_{\mathbf{X}},\Pi)} G_{k_n}\left(\hat{Q}^{2k_n+1}[Z^n],\Pi,f\right)\right] - \max_{(P_{\mathbf{X}},\Pi)} G_{k_n}\left(\hat{Q}^{2k_n+1}[Z^n],\Pi,f\right)\right| = 0 \quad P_{\mathbf{Z}}-a.s..$$

The definition of $G_k$, via an elementary continuity argument, is readily verified to imply

$$\left|E_{\mathbf{Z}}\left[\max_{(P_{\mathbf{X}},\Pi)} G_{k_n}\left(\hat{Q}^{2k_n+1}[Z^n],\Pi,f\right)\right] - \max_{(P_{\mathbf{X}},\Pi)} G_{k_n}\left(\hat{Q}^{2k_n+1}[Z^n],\Pi,f\right)\right| \leq$$

$$\max_{\Pi\in\Delta}||\Pi^{-1}||\Lambda_{\max}|\mathcal{A}|^{6k_n+3}\left|\left|E_{\mathbf{Z}}\left[\hat{Q}^{2k_n+1}[Z^n]\right] - \hat{Q}^{2k_n+1}[Z^n]\right|\right|.$$

By the construction of $\alpha_\psi$, for any $\varepsilon > 0$ we have

$$P_{\mathbf{Z}}\left(\left|E_{\mathbf{Z}}\left[\max_{(P_{\mathbf{X}},\Pi)} G_{k_n}\left(\hat{Q}^{2k_n+1}[Z^n],\Pi,f\right)\right] - \max_{(P_{\mathbf{X}},\Pi)} G_{k_n}\left(\hat{Q}^{2k_n+1}[Z^n],\Pi,f\right)\right| > \varepsilon\right) \leq$$

$$\alpha_\psi\left(n, k_n, \frac{\varepsilon}{\max_{\Pi\in\Delta}||\Pi^{-1}||\Lambda_{\max}|\mathcal{A}|^{6k_n+3}}\right). \tag{94}$$

Since by hypothesis the right hand side is summable, by the Borel–Cantelli lemma

$$P_{\mathbf{Z}}\left(\limsup_{n\to\infty}\left|E_{\mathbf{Z}}\left[\max_{(P_{\mathbf{X}},\Pi)} G_{k_n}\left(\hat{Q}^{2k_n+1}[Z^n],\Pi,f\right)\right] - \max_{(P_{\mathbf{X}},\Pi)} G_{k_n}\left(\hat{Q}^{2k_n+1}[Z^n],\Pi,f\right)\right| > \varepsilon\right) = 0.$$

Since $\varepsilon$ is arbitrary, we can take $\varepsilon \to 0$ and get

$$\lim_{n\to\infty}\left|E_{\mathbf{Z}}\left[\max_{(P_{\mathbf{X}},\Pi)} G_{k_n}\left(\hat{Q}^{2k_n+1}[Z^n],\Pi,f\right)\right] - \max_{(P_{\mathbf{X}},\Pi)} G_{k_n}\left(\hat{Q}^{2k_n+1}[Z^n],\Pi,f\right)\right| = 0 \quad P_{\mathbf{Z}}-a.s.$$

The proof is now completed similarly to the proof of Claim 1. □

Equipped with Claim 2, we now complete the proof of Proposition 1 as follows. We have

$$\limsup_{n\to\infty}|B_n| \leq \limsup_{n\to\infty}\left|\min_{f\in\mathcal{F}_{k_n}}\max_{(P_{\mathbf{X}},\Pi)} E_{\mathbf{Z}}[g_{P_{\mathbf{X}},\Pi,f}(\mathbf{Z})] - \min_{f\in\mathcal{F}_{k_n}}\max_{(P_{\mathbf{X}},\Pi)} G_{k_n}\left(\hat{Q}^{2k_n+1}[Z^n],\Pi,f\right)\right|$$

$$+ \limsup_{n\to\infty}\left|\min_{f\in\mathcal{F}_{k_n}} E_{\mathbf{Z}}\left[\max_{(P_{\mathbf{X}},\Pi)} G_{k_n}\left(\hat{Q}^{2k_n+1}[Z^n],\Pi,f\right)\right] - \min_{f\in\mathcal{F}_{k_n}} E_{\mathbf{Z}}\left[\max_{(P_{\mathbf{X}},\Pi)} g_{P_{\mathbf{X}},\Pi,f}(\mathbf{Z})\right]\right|$$

$$+ \limsup_{n\to\infty}\left|\min_{f\in\mathcal{F}_{k_n}} E_{\mathbf{Z}}\left[\max_{(P_{\mathbf{X}},\Pi)} G_{k_n}\left(\hat{Q}^{2k_n+1}[Z^n],\Pi,f\right)\right] - \min_{f\in\mathcal{F}_{k_n}}\max_{(P_{\mathbf{X}},\Pi)} G_{k_n}\left(\hat{Q}^{2k_n+1}[Z^n],\Pi,f\right)\right|.$$

From (92), (93), Claim 2, and the fact that $|B_n| \geq 0$ it follows that

$$\lim_{n\to\infty}|B_n| = 0 \quad P_{\mathbf{Z}}-a.s. \tag{95}$$

Combined with (95) and (89) this gives

$$\limsup_{n\to\infty}\left[\max_{(P_{\mathbf{X}},\Pi)} E_{[P_{\mathbf{X}},\Pi]}\left[L_{\hat{X}^n_{univ}}(X^n,Z^n)\right] - \min_{f\in\mathcal{F}_{k_n}}\max_{(P_{\mathbf{X}},\Pi)} E_{[P_{\mathbf{X}},\Pi]}\left[L_{f\in\mathcal{F}_{k_n}}(X^n,Z^n)\right]\right] \leq 0 \quad . \tag{96}$$

On the other hand, since $\hat{X}^n_{univ} \in \mathcal{F}_{k_n}$,

$$\max_{(P_{\mathbf{X}},\Pi)} E_{[P_{\mathbf{X}},\Pi]} \left[ L_{\hat{X}^n_{univ}}(X^n, Z^n) \right] \geq \min_{f \in \mathcal{F}_{k_n}} \max_{(P_{\mathbf{X}},\Pi)} E_{[P_{\mathbf{X}},\Pi]} \left[ L_{f \in \mathcal{F}_{k_n}}(X^n, Z^n) \right].$$

When combined with (96), we get the desired result

$$\lim_{n \to \infty} \left| \max_{(P_{\mathbf{X}},\Pi)} E_{[P_{\mathbf{X}},\Pi]} \left[ L_{\hat{X}^n_{univ}}(X^n, Z^n) \right] - \min_{f \in \mathcal{F}_{k_n}} \max_{(P_{\mathbf{X}},\Pi)} E_{[P_{\mathbf{X}},\Pi]} \left[ L_{f \in \mathcal{F}_{k_n}}(X^n, Z^n) \right] \right| = 0.$$

□